\documentclass[conference]{IEEEtran}

\usepackage[utf8]{inputenc}
\usepackage{amsfonts}
\usepackage{amsmath}
\usepackage{graphicx}
\usepackage{algorithm2e}
\usepackage{balance}
\usepackage[group-separator={,}]{siunitx}
\usepackage{epstopdf}
\graphicspath{{../}}

\newtheorem{lemma}{Lemma}
\newtheorem{theorem}{Theorem}

\newtheorem{remark}{Remark}
\newtheorem{claim}{Claim}

\newcommand{\boldheader}[1]{\vskip 5pt \noindent{\bf #1}}
\newcommand{\alg}{\mathcal{A}}
\newcommand{\cent}{\mathcal{C}}

\title{A DoS Attack on Off-Chain Networks\\and A Design Tradeoff}

\title{DoS Attack on Off-Chain Networks:\\Feasibility and Design Tradeoffs}

\title{The Price of Predictability:\\A Novel DoS Attack on Off-Chain Networks}

\title{The Price of Predictability:\\Route Hijacking and DoS in Off-Chain Networks}

\title{Hijacking Routes in Payment Channel Networks:\\A Predictability Tradeoff}

\author{
\IEEEauthorblockN{Saar Tochner and Aviv Zohar}
\IEEEauthorblockA{The Hebrew University of Jerusalem\\
\{saart,avivz\}@cs.huji.ac.il}
\and
\IEEEauthorblockN{Stefan Schmid}
\IEEEauthorblockA{Faculty of Computer Science, University of Vienna\\
stefan\_schmid@univie.ac.at}}
\begin{document}

\maketitle

\begin{abstract}
Off-chain transaction networks can mitigate the scalability
issues of today's trustless electronic cash systems such as Bitcoin. 
However, these peer-to-peer networks also introduce a new attack surface which is not well-understood today.
This paper identifies and analyzes,  a novel Denial-of-Service attack
which is based on route hijacking, i.e., 
which exploits the way transactions are routed and executed along 
the created channels of the network.
This attack is conceptually interesting as even a limited attacker that manipulates the topology through the creation of new channels
can navigate tradeoffs related to the way it attacks the network. 
Furthermore, the attack also highlights a fundamental design tradeoff for the defender
(who determines its own routes): 
to become less predictable and hence secure, a rational node has to pay higher fees to nodes that forward its payments.
We find that the three most common implementations for payment channels in Bitcoin (lnd, C-lightning, Eclair) approach routing differently. We begin by surveying the current state of the Lightning network and explore the routes chosen by these implementations. We find that in the current network nearly 60\% of all routes pass through only five nodes, while 80\% go through only 10 nodes. Thus, a relatively small number of colluding nodes can deny service to a large fraction of the network. 

We then turn to study an external attacker who creates links to the network and draws more routes through its nodes by asking for lower fees. We find that just five new links are enough to draw the majority (65\% - 75\%) of the traffic regardless of the implementation being used.  The cost of creating these links is very low.

We discuss the differences between implementations and eventually derive our own suggested routing policy, which is based on a novel combination of existing approaches. 
\end{abstract}

\section{Introduction}\label{sec:intro}
Emerging decentralized ledger and blockchain technologies bear the promise to
streamline business, governance and non-profit activities, by
eliminating intermediaries and authorities.
A main hurdle toward such more decentralized applications however
remains scalability~\cite{bamert2013have,decker2015fast,sompolinsky2013accelerating}. The typical example is that Bitcoin can only support
dozens of transactions per second, compared to several thousands
in deployed payment services such as Visa~\cite{trillo2013stress}. 

Off-chain peer-to-peer networks (a.k.a.~payment channel networks) 
are a promising approach to mitigate
this scalability problem: by allowing participants to 
make payments directly through a network of payment channels,
the overhead of global consensus protocols and committing transactions \emph{on-chain} 
can be avoided.
This not only improves transaction throughput
but also avoids the blockchain's transaction latency;
ideally, in a payment channel network, most transactions are done using bidirectional payment
channels that only require direct communications between a handful of nodes, while the blockchain is used only rarely, to establish or terminate channels. 
As an incentive to participate in others' transactions, the nodes obtain a small fee from every transaction that was routed through their channels.
Over the last few years, payment channel networks such as Lightning~\cite{poon2016bitcoin}, Ripple~\cite{armknecht2015ripple}, and Raiden~\cite{network2018cheap}
have been implemented, deployed and have started growing.

This paper is concerned with the \emph{routing} mechanisms which lie at the heart
of payment channel networks:
an important feature of payment channel networks is that they
also support transactions between participants without direct channels, 
using \emph{multihop} routing~\cite{malavolta2019anonymous,poon2016bitcoin}.
However, the design tradeoffs and security implications of such multi-hop routing
are not well-understood today.
In fact, routing in payment channel networks is 
fairly different from routing in traditional communication networks: 
in traditional communication networks, routing algorithms typically
aim to find short and low-load paths in a network whose links are subject
to fixed capacity constraints. In a payment channel network, 
link capacities represent
payment balances, which can be highly dynamic: every transaction changes the payment balance initially
set up for the channel. 
What is more, both the establishment as well as the use of payment
channels is an inherently strategic decision, and subject to complex
incentives and the extent to which a participant thinks she or he can benefit from
different behaviors. 
In fact, a participant may not only try to strategically maximize her or his
profit, but may also be \emph{malicious}.

\subsection{Our Contributions}

This paper is motivated by the question whether and how malicious players
can strategically influence and \emph{exploit} the way transactions are routed
in off-chain networks. 
Our main contribution is the identification, analysis, and evaluation of a novel
Denial-of-Service attack which is based on the
hijacking of transaction routes. To this end,
we examine different existing
implementations (which turn out to differ significantly),
and we provide empirical insights into the structure and properties
of payment channel networks. 

More specifically, we find that there exists a group of 10 nodes that participates in 80\% of the routes, and 30 nodes that participate in more than 95\% of the routes. On the other hand, we find that by creating 5 new \emph{new} channels, an attacker can hijack about 65\% of the routes, and with 30 channels, it can hijack 80\% of the routes of every implementation.
Furthermore, we find that existing clients differ in their
objectives and hence introduce trade-offs in route selection. Finally, we find only 
limited evidence that users configure their nodes to extract high rewards. Nodes typically use default values or set minimal fees and contribute cheap routes to the network.
Both aspects can potentially be exploited by selfish and malicious
players. 

Reasoning about more secure solutions, we find that the underlying
routing problem exhibits a fundamental design tradeoff, related to how
unpredictable a rational player aims to be:
by behaving less predictably (i.e., choosing alternative routings), 
a rational player can become more secure
against attacks, however, such behaviors induce a higher fee for the player. 
We investigate this novel tradeoff (the price of predictability) and discuss strategies both for
the attacker as well as the defender to optimally invest its resources. 

An additional tradeoff that we discuss 
regards the times nodes must be online 
(how often they need to check the network to make sure funds are not stolen from them) 
and the time it takes live channels to reset. Network defaults currently allow nodes relatively long periods of
offline time, which implies that connection attempts that are hijacked are not retried quickly.

This paper investigates additional tradeoffs and gives intuition on their impact on the network. One of these tradeoffs regards the fee rate of the channels. On the one hand, if we  let the nodes determine the fee by themselves (in an open market), then it will increase the vulnerability to attacks in which the attacker will exploit the nodes' selfish routing and offer very low fees in order to hijack traffic. On the other hand, if the system  determines the fees for the channels, then the system may suffer from incentives problems; too low fees will cause the channels to be non-profitable, which may cause them to close (which affects the network connectivity); too high fees will decrease the incentives of the nodes to perform transactions.

In order to ensure reproducibility and in order to facilitate followup work,
we share our source code (python3.7) with the research community at https://github.cs.huji.ac.il/saart/saart-lightning.

\subsection{Organization}

The remainder of this paper is organized as follows.
In Section~\ref{sec:attack}, we will present our DoS attack
and explore different hijack possibilities.
Section~\ref{sec::feasibility} examines the state-of-the-art
routing algorithms, and then provides an overview of the experiments we conducted. We then describe in Section~\ref{sec:preliminaries} a model resulting from our attack
as well as the algorithms that we used in our experiments. In Section~\ref{sec::suggested_solution} we explore methods to decrease this vulnerability through both a game theoretic model and suggestions on future weight implementation. Related work will be presented in
Section~\ref{sec:relwork}.
Finally, we present conclusions and discuss future work in Section~\ref{sec:conclusion}.

\section{DoS Attack Via Route Hijacking}\label{sec:attack}

This section uncovers a potential vulnerability, based on route hijacking,
which may be used for
a Denial-of-Service (DoS) attack on off-chain networks.
We will first provide an explanation of the key elements of the protocol,
then present the basic attack and finally
describe how this attack may be amplified.
In the next sections, we will then explore the feasibility of this attack empirically,
in a state-of-the-art payment channel network, 
evaluate it and reason about optimization opportunities both for the attacker
and the defender.

\subsection{Context of the Attack}
To be more concrete, we consider the Lightning network
as a case study in the following. However, the concepts are similar
in other networks as well. 
In the Lightning off-chain network, the \emph{\textbf{channels}} are established
by the nodes for secure payments. 
Every two nodes that are willing to create such a channel,
need to make a commitment: they need to execute a Bitcoin transaction that locks
money (i.e., liquidity) for this channel\footnote{Note that this means that every channel is backed-up with a real Bitcoin's transaction, therefore no one can spoof channels}.
A transaction is then simply an agreement between the two end-points
of the channel, which leads to a different split of the money. 
The intermediate states resulting from this transaction
do not have to be committed to the blockchain: Once they will commit the state into the blockchain, the channel will be closed (because it ``wastes" the original transaction). Until this occurs, the channel can remain operational and the internal split of funds can be adjusted by the participants. 
As the intermediate states of channels are built, older states are ``revoked'': if one tries to commit an old state, the other participant can claim funds back. This recovery of funds can only be done within a certain pre-set period of time. This setup thus requires each of the participants in a channel to occasionally check the blockchain and make sure that the other party did not close the channel using an old state. 

Off-chain networks such as Lightning however do not only support transactions between
nodes that have a direct channel, but also allow chaining paths to connect nodes \emph{indirectly}. 
This is achieved by allowing two nodes to find a path along multiple existing channels that 
connect the nodes: in order to realize such a ``multi-hop transaction'',
a transaction is executed on each direct channel along the path. 

The technique used to chain paths together but still guarantee that funds are not stolen by intermediate nodes is based on ``hash \& time lock contracts", or HTLCs, which are essentially contracts awarding nodes a slightly different split of the money in each channel if a secret is revealed. Paths are then created by establishing a chain of channels with HTLCs conditioned on the release of the same secret, and the transfer is finally executed as the recipient node releases the secret (additional details can be found in~\cite{poon2016bitcoin}).
HTLCs must additionally posses an expiration time which specifies the timeout of each lock. This timeout is the timeout of the next node in the path plus some small delay specified by the preceding node. These decreasing delays ensure that intermediate nodes never reach a situation where they might have an outgoing payment without being compensated by an incoming payment (due to an earlier timeout of the incoming channel). 
This small difference between HTLC timeouts is called \emph{\textbf{delay}} and is expressed by a number of blocks in the blockchain (as timestamps on blocks are considered unreliable, and block height is the fundamental way to meassure the progress of time in blockchains).

To motivate nodes to allow routing transactions, nodes are allowed to specify a  \emph{\textbf{fee}} for forwarding transactions.
This fee that the nodes performing a transaction have to pay to the nodes that hold the channels that they use,
is published when the channel is created. It consists of a \emph{base fee} 
and a \emph{proportional fee}; the latter is relative to the transaction size. 
For example, to use a channel that has base fee of 100 millisatoshis, and a proportional fee of 1 per million, 
a 1 million millisatoshis transaction will pay a fee of 101, and a 
2 million millisatoshis transaction will pay a fee of 102.

In order to find such a path, nodes leverage the knowledge of the channel graph which is
continuously gossiped about by nodes in the network. 
Given this knowledge of the network graph, nodes utilize \emph{source routing} to pick their path. As we will see, different implementations use different 
routing algorithms for path selection, optimizing different measures (e.g. fee, delay, security, etc.).

\subsection{Basic Attack: A Rerouting Vulnerability}

    \begin{figure}
    	\centering
    	\includegraphics[scale=0.22]{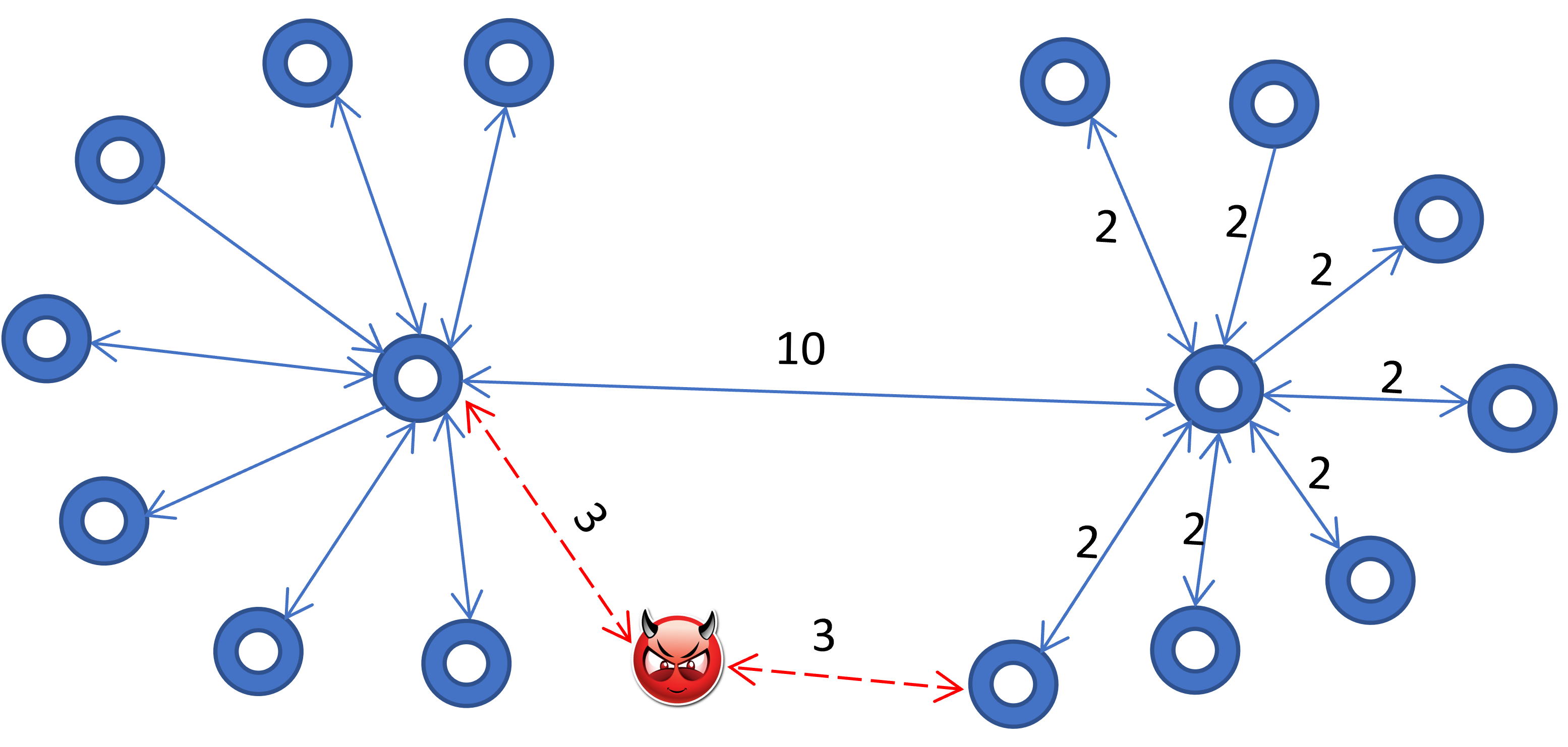}
    	\caption{The routing vulnerability: an adversary creates edges that decrease the weight for many nodes.}
    	\label{figure::attack_ilustration}
    \end{figure}

The fact that nodes can strategically choose transaction paths 
introduces a potential vulnerability. In the following,
we model the off-chain network as a graph, 
where nodes represent the Lightning nodes of the network and 
edges represent payment channels. 
In the basic attack, an 
adversary (either selfish or malicious) can aim to establish a set of edges 
in this graph which put it in a topologically important location,
as well as to announce a low fee.
As a consequence, other nodes are likely to route transactions
through the adversarial node. As route establishment is done via an onion-routing approach, and intermediary node may drop the payload and fail to follow through on the establishment of the rest of the path. In this case, the payment does not take place, and the sender must wait for the original HTLC to expire before attempting to re-send the payment. 

By maximizing its \emph{centrality}, the adversary can
hijack a large number of transaction paths, which in turn allows it to launch a Denial-of-Service
attack. Even if payments are re-sent, route selection may yet again cause the path to go through one of the attacker's nodes. 

For an illustration, consider Figure~\ref{figure::attack_ilustration}.
First, ignore the red node and its edges. In the blue network, 
there are then two groups of nodes that communicate
through a single link with a high fee of 10. 
If an adversary (indicated as red node) now introduces 
the two red edges of low fee, it essentially creates 
a shortcut between some of the nodes, hijacking 
the transactions that aim to minimize the fee.

While we will analyze the algorithmic problem underlying
both the adversary's and the defender's optimization more formally only later,
we note that counteracting this attack is non-trivial.
Essentially, there are two options:
\begin{enumerate}
    \item  One may consider introducing mechanisms
    which quickly alert nodes about interrupted channels.
    However, this can also be problematic as it may be exploited
    by adversaries to make false reports.
    \item If the source node does not know which specific channels were stopped, 
    it may only heuristically invalidate nodes or channels from the original path 
    (which may disconnect the network or lead to higher fees), 
    and/or hope that a newly (randomly?) chosen path may reestablish
    connectivity.
\end{enumerate}

Note that every node can create channels to most of the nodes that it chooses:
the default behaviour in all the implementations is to accept every channel suggestions. This willingness to connect can be attributed to the perceived low risk in doing so: the construction guarantees that none of the funds in the channel is at risk of theft. Given the attack that we propose, it may be wise to accept channel connections only from known and trusted entities.

\subsection{Amplified Attack: Delay Vulnerability}

The basic attack may be further amplified
by exploiting the delay mechanism. 
    If the attacker participates in path establishment, 
    but stops participating during the transaction itself, then the other 
    nodes may already have locked their money for the channel
    and will be able to free it only after a timeout.
    This means that a higher delay will lock the money of the nodes in the path for a longer time, 
    and potentially prevent from the source node to execute another transaction's attempt for a longer period. We explore this aspect as well in our evaluation. 
    
\section{Feasibility and Case Studies}\label{sec::feasibility}

We now explore the feasibility of the attack identified
above. To this end, we consider different Lightning network implementations
and also conduct an empirical study on today's
network topology, its fees and other parameters, which may be of independent interest.
We also report on our experimental evaluation results.

\subsection{Implementation Details} \label{sec::imp_details}

In order to investigate the feasibility of our attack,
as a case study, we consider the three 
main implementations of the Lightning network: 
\emph{lnd} (implemented in Golang), \emph{C-lightning} (implemented in C) and 
\emph{Eclair} (implemented in Scala).
The implementations differ in the way they operate relative to aspects not covered
in the BOLTs~\cite{BOLT} which make up the Lightning network's standard. Specifically, the standard does not dictate any routing behavior, leaving each implementation to set its own (as there is no real need for path finding to be identical between implementations). In our experiments, we use the default parameters of every implementation.

\subsubsection{lnd}
\emph{lnd} chooses the path of minimum weight, 
calculated using the following recursive formula, 
where $p$ is the list of channels in this path, 
and $ams$ is the list of amounts the go
through each channel (changes depending on the fees):\footnote{References can be found in the methods \emph{FindRoutes} and \emph{findPath} in lnd.routing.router.go and pathfind.go}
$$ fee = ams[i+1] \cdot p[i].propFee + p[i].baseFee $$
\begin{align*}
    weight[i] = &ams[i+1] \cdot p[i].delay \cdot riskFactor \\
                &+ fee
\end{align*}
For default $riskFactor = 15 / \num{1000000000}$

Note that lnd changed this weight function in March~2019, in commit 6b70791. The authors added\footnote{References can be found in routing/missioncontrol.go:261 and routing/pathfind.go:531} a new phrase to the channel's weight: $+ \frac{100}{edgeProbability}$.
This parameter is an aggregated success score over the previous routing through this channel. If the node has no prior knowledge about the channel, then it uses the default value that is relative to the a-priori failing rate in the network. Otherwise, the penalty considers only the time of the last failure. In the first hour, the probability is $0$, and then it increases exponentially with the formula: $0.6 - \frac{0.6}{2^t}$ (when using the default arguments).
%
%The authors added to the weight function the amount %$$\frac{PaymentAttemptPenalty}{edgeProbability}$$
%for a default $PaymentAttemptPenalty = 100$. If there is no history, this probability is %$AprioriHopProbability=0.6$. If the last route was successful, then the probability jumps to %$0.95$; and otherwise, only the time of the last fail $t$ is examined (in number of hours ago, %rounded downward), and the probability will be $AprioriHopProbability \cdot \big(1 - %2^{-t/PenaltyHalfLife}\big)$, where $PenaltyHalfLife=1$ by default. I.e., the default %probability after a fail is $0.6 - \frac{0.6}{2^t}$.
%
We note that lnd looks only at the last failure to discount channels, so if the time until failure is long enough, then it will effectively choose between at most two channels in the network, both with low weight (quite similarly to Eclair's ``top 3" approach). This is not in itself sufficient to bypass the attacker, as it is easy for it to be on both routes (just like we show for Eclair). Additionally, the decay rate of the penalty on failures needs to be low in order to remain relevant as the previous HTLC contract times out. At last, an attacker that completes the route as requested, but delays the HTLC secret release until the very last moment, will delay the transfer significantly and will not suffer the penalty at all.

\subsubsection{C-lightning}

\emph{C-lightning} multiplies 
the fee by a fuzz that is 
randomly calculated (and is within configured range), and gives a
penalty for delays.
Denote by $h$ the hash that was calculated using siphash24 
on a random string that the user generated (before every path selection) and the short channel id. Denote $fuzz$ to be the configured fuzz factor ($0.05$ by default).\footnote{References can be found in the methods bfg\_one\_edge and find\_route in the file gossipd/routing.c}
$$ scale = 1 + fuzz \cdot (2 \cdot \frac{h}{2^{64}-1} - 1) $$
$$ fee = scale \cdot (ams[i+1] \cdot p[i].propFee + p[i].baseFee) $$
\begin{align*}
    weight[i] = &(ams[i+1] + fee) \cdot (p[i].delay \cdot riskFactor) \\
                &+ 1
\end{align*}
for a configurable $riskFactor$, which is 10 by default.

\subsubsection{Eclair} \label{subsubsec::imp_eclair}

\emph{Eclair} multiplies 
the fee by a proportional factor depending on the channel properties: 
delay, capacity, and height (while assuming upper and lower bound for each of them).
In addition to the above, Eclair also randomizes 
the selected paths uniformly, from the 3 (a parameter) best routes.\footnote{References can be found in: eclair-core/src/main/resources/reference.conf, and the methods FindRoute, edgeWeight in the files eclair/router/Router.scala, Graph.scala}

$$ fee = ams[i+1] \cdot p[i].propFee + p[i].baseFee $$
\begin{align*}
 weight[i] = fee \cdot &(normalizedDelay \cdot delayRatio \\
 &+ normalizedCapacity \cdot capacityRatio \\
 &+ normalizedHeight \cdot ageRatio) 
\end{align*}
For upper and lower bounds\footnote{8640 is the number of Bitcoin's blocks in two months}: 
\begin{align*}
    9 &< delay < \num{2016} \quad delayRatio = 0.15 \\
    \num{1000} &< capacity < 2 ^{24} \quad capacityRatio = 0.5 \\
    0 &< height < \num{8640} \quad ageRatio = 0.35
\end{align*}

\subsection{First Empirical Insights}

We first provide some general insights into the empirical
properties of today's Lightning network.
While these insights are not directly related to 
routing, they provide insights into the behavior
of the network users, and what this implies
for the vulnerability of the network.

\subsubsection{Methodology}

The following results are based on measurement data we collected using 
a live Lightning node (lnd) 
that is connected to the mainnet through bitcoind.
We used the CLI command $lncli~ describegraph$ in order to extract the network structure,
and mongodb to store and query the different paths (different implementations, and different parameters).
%The graphs were plotted with $matplotlib$.
We use information that was dumped from a live mainnet Lightning node at July 24th 2019 17:00 UTC.

Note that using this method, we can examine only public channels;
our analysis omits private channels.
We argue that since these channels are private,  nodes which are not directly part of the private channel are typically not aware of them and will not route through them. 
Thus most routing, by design, relies primarily on the public network.

\subsubsection{Network Analysis}

    The network is composed of
    4,300 nodes, 33,600 channels, 
    with an average channel capacity of 0.028BTC and an average node capacity of 0.238BTC.

\begin{figure}
    	\centering
    	\includegraphics[scale=0.6]{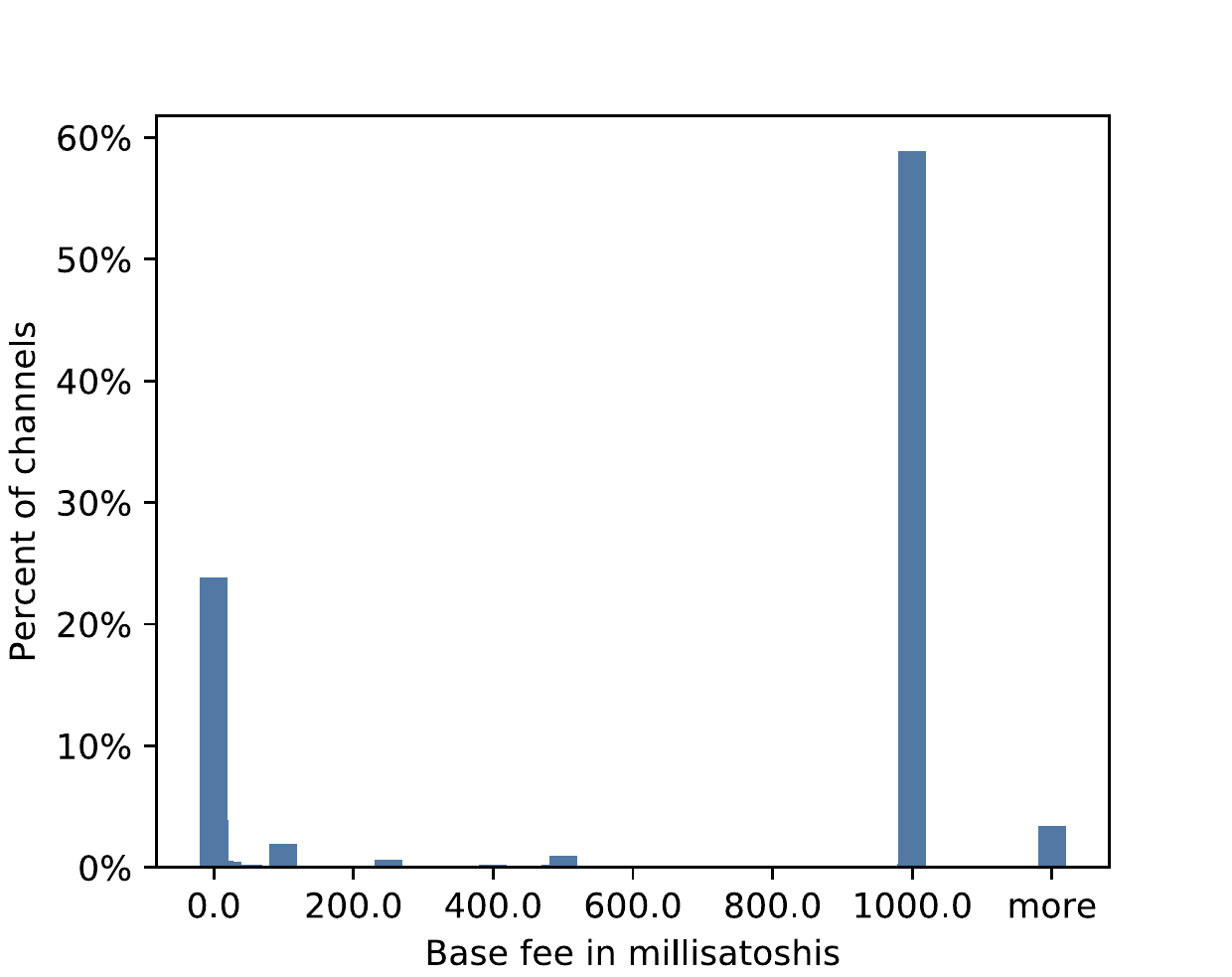}
    	\caption{Channels by base fees}
    	\label{figure::base_fee}
    \end{figure}

\begin{figure}
    	\centering
    	\includegraphics[scale=0.6]{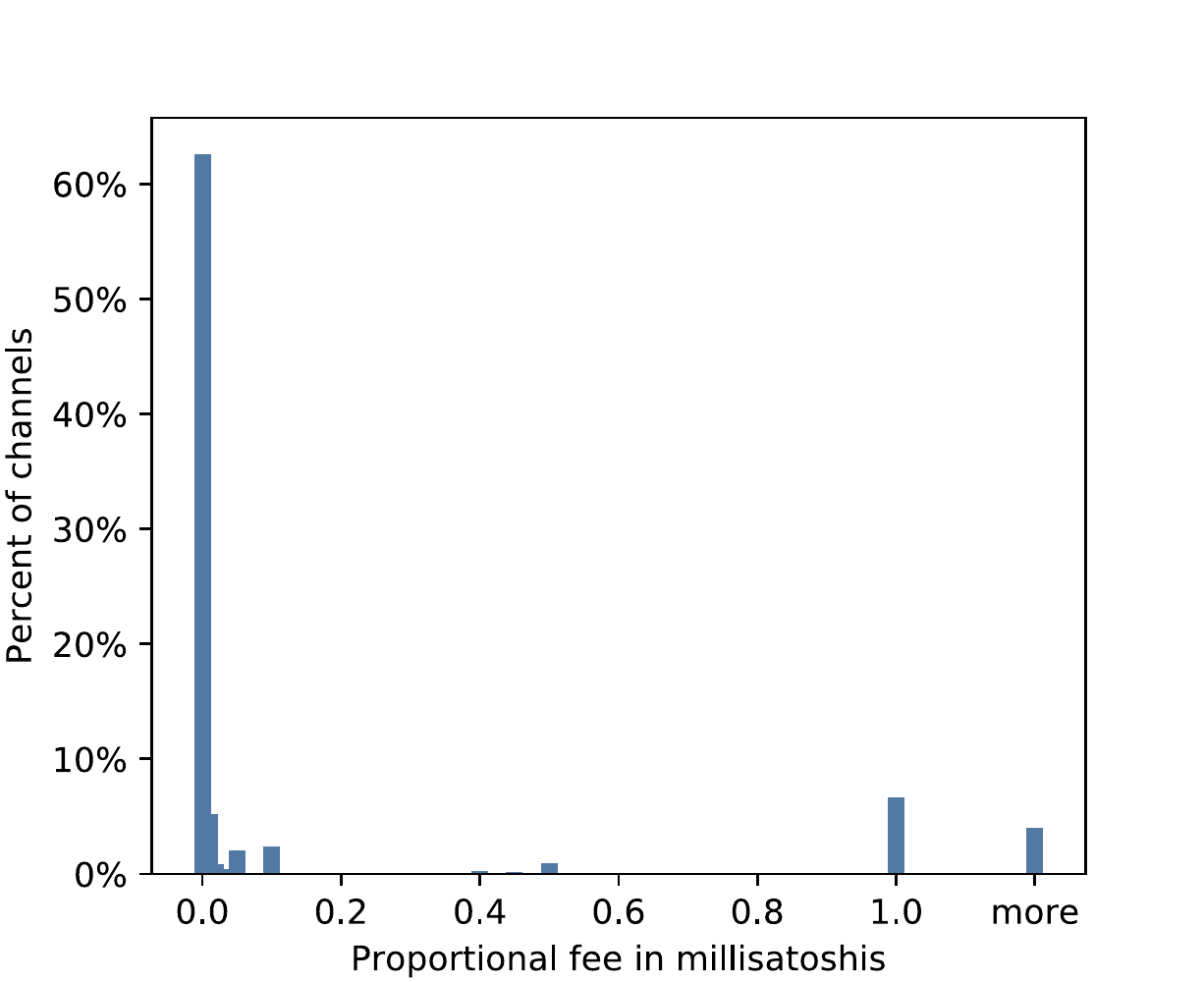}
    	\caption{Channels by proportional fees}
    	\label{figure::prop_fee}
    \end{figure}

\begin{figure} 
    	\centering
    	\includegraphics[scale=0.6]{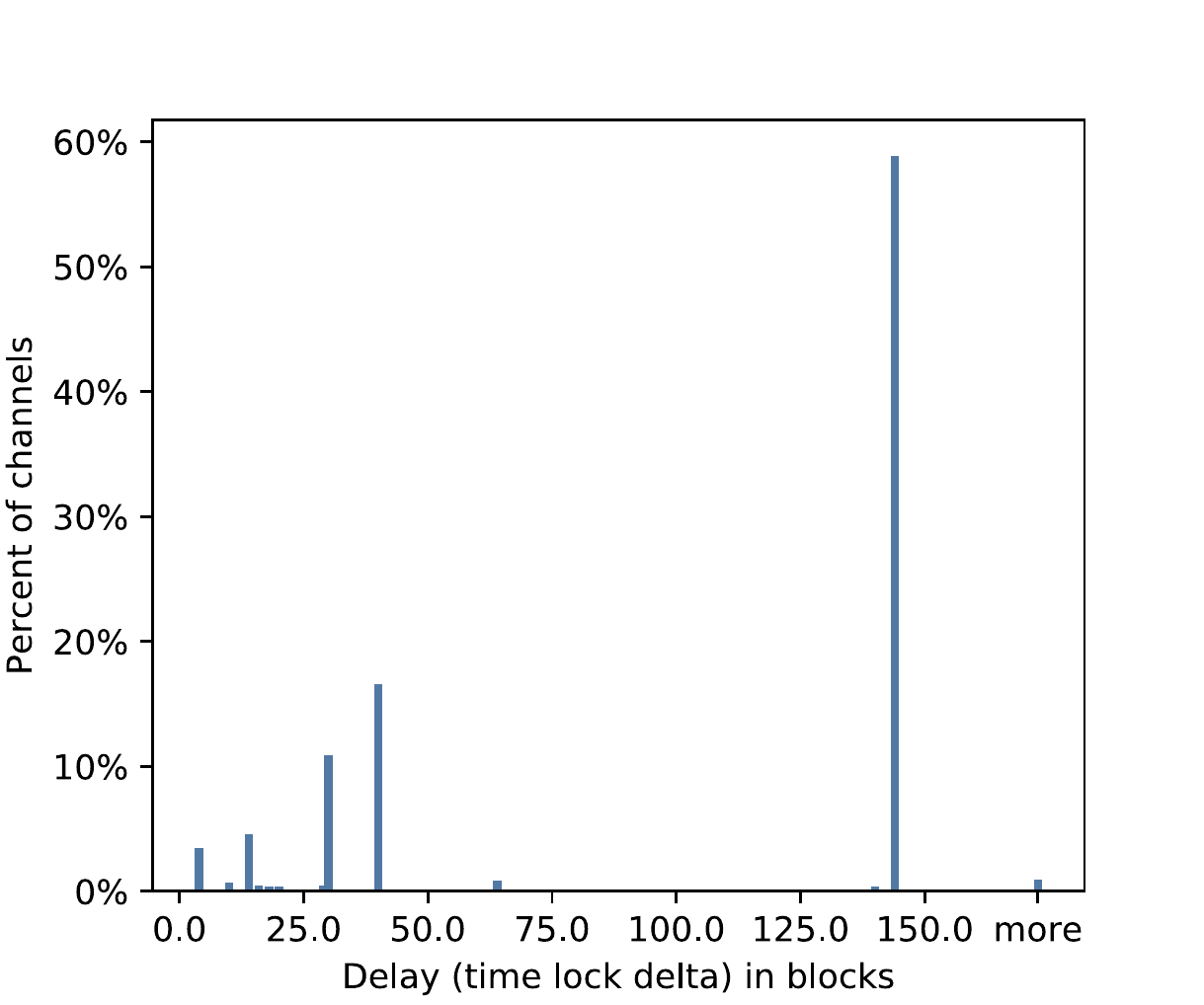}
    	\caption{The delay of the channels (144 blocks is $\sim 24$ hours)}
    	\label{figure::delay}
    \end{figure}

\begin{figure}
    	\centering
    	\includegraphics[scale=0.6]{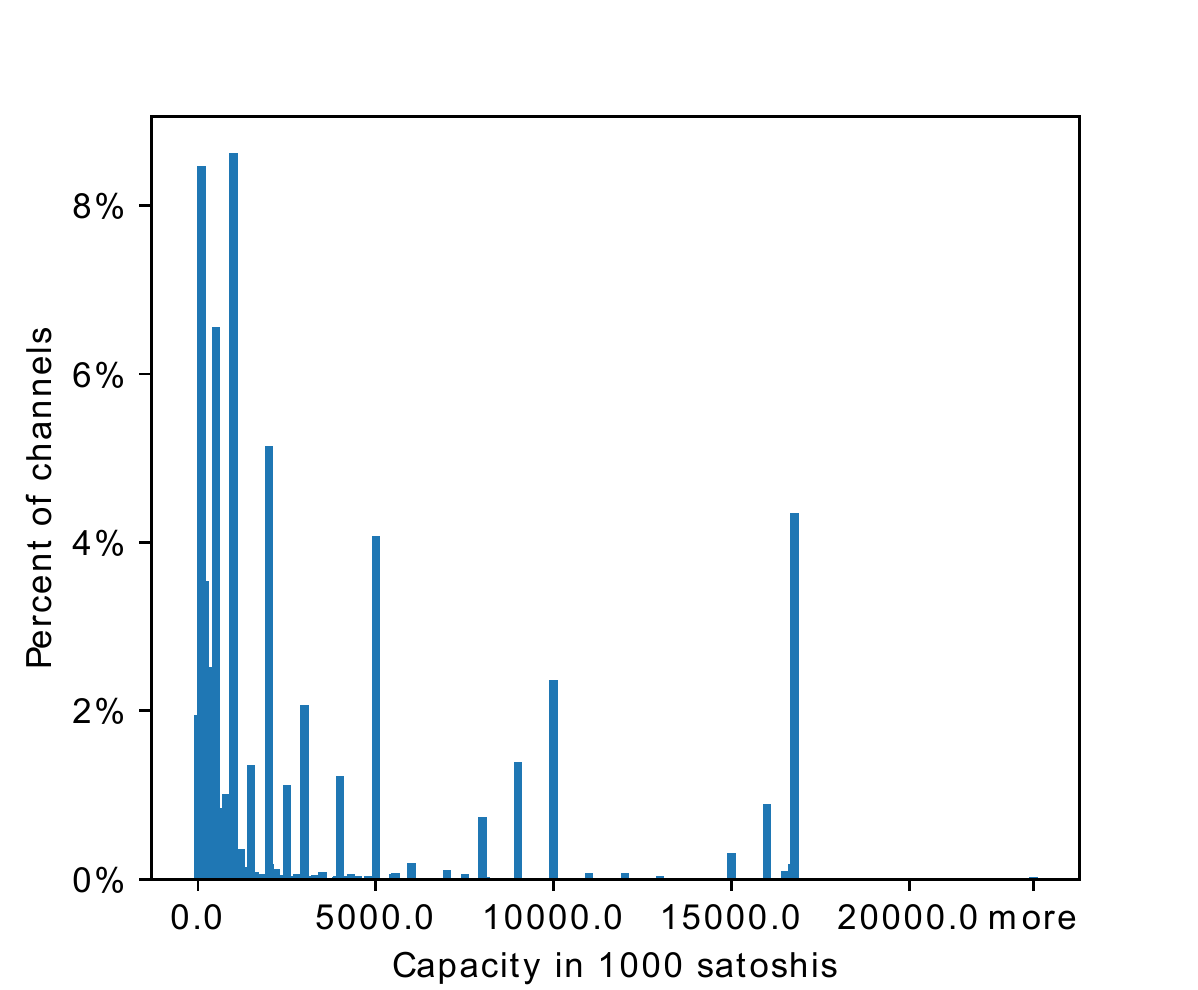}
    	\caption{Channels by capacities}
    	\label{figure::capacity}
    \end{figure}

\begin{figure}
    	\centering
    	\includegraphics[scale=0.6]{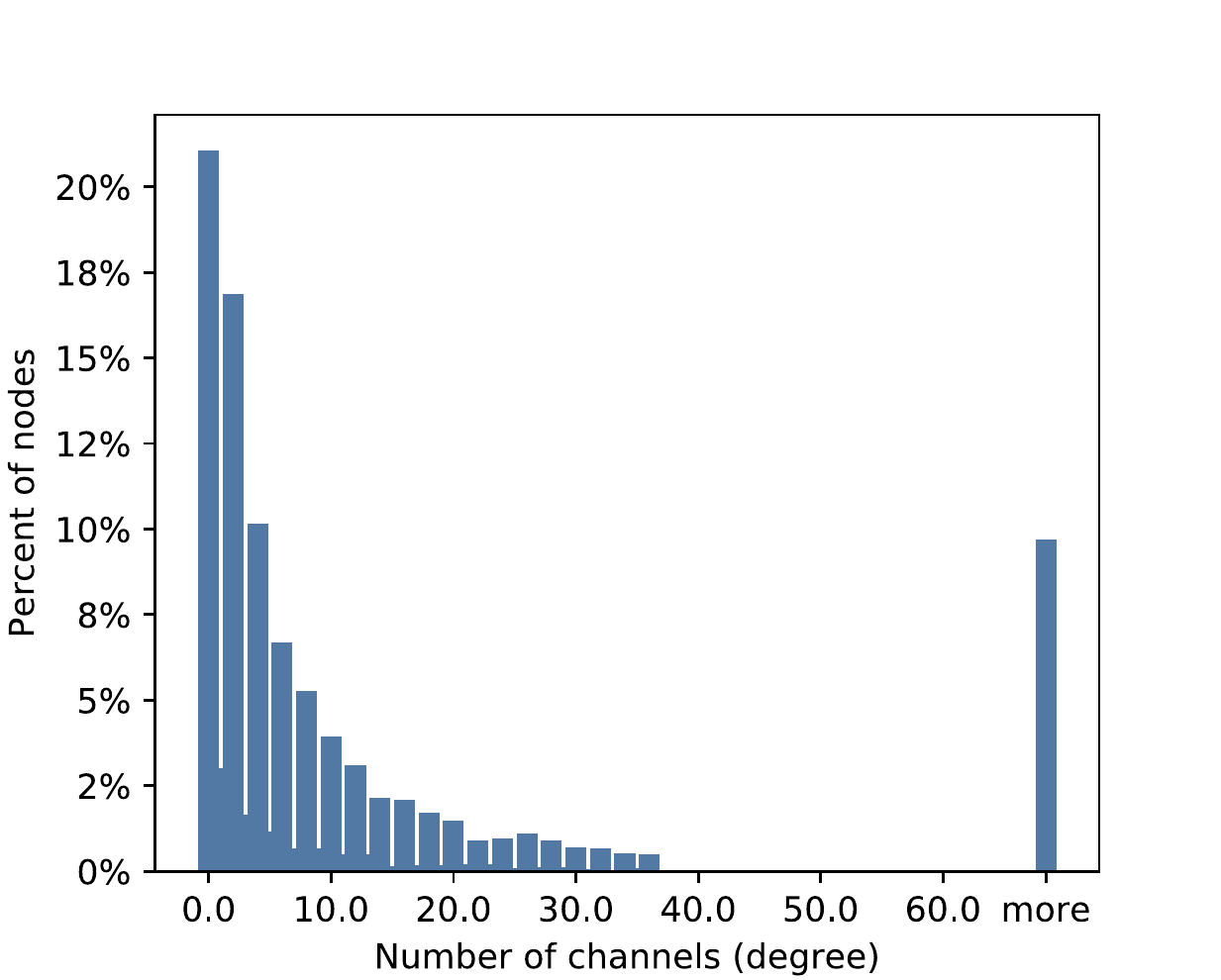}
    	\caption{number of channels per node}
    	\label{figure::degree}
    \end{figure}

    Figures~\ref{figure::base_fee}, \ref{figure::prop_fee}, \ref{figure::delay}, \ref{figure::capacity}, and \ref{figure::degree} show some basic properties on the Lightning network.
    In particular, Figure~\ref{figure::base_fee} reveals that the base fee across channels has two highly common values:
    most channels simply use the default value, which is 1000.
    Interestingly, however, the second most frequent value,
    and the most frequent non-default value, is the minimum possible fee.
    This provides two main insights: first,  most
    users do not configure the software beyond the default values,
    and second, most of those who do, do it in a way which \emph{supports} the network.
    Thus, we hardly find any evidence for selfish optimizations of fees in the current network.
    Both properties may influence a potential attacker.
    
    Figure~\ref{figure::prop_fee} shows the corresponding distribution
    for the proportional fee, the transaction fee.
    Here, the default value is 1/1000, which is also by far the most frequent value.
    Interestingly, however, the value 1 is also frequent; we conjecture that this may be due
    to a confusion with the base fee, or with units (satoshis vs millisatoshis). 
    Other high values also appear, which one may interpret as an attempt to profit from the network---but this is unlikely: the values are still very small. Given the network scale,
    nodes are unlikely to be able to benefit from such fees~\cite{lightning2018paylittle}. This figure also suggests that there are altruistic nodes in the network, which are willing to hold channels without taking fees. Indeed, out of the channels with base fee 0, we obtain that the percentage of channels with 0 proportional fee is about twice the percentage in the whole network (about 40\% comparing to 24\%).
    
    Similarly to the other figures, Figure~\ref{figure::delay} shows that most of the channels in the network use the defaults. In this case, we see that 144 blocks (a full day), are used as delay by most of the channels. Note that this percentage is similar to the percentage of the default configuration in Figures~\ref{figure::base_fee} and \ref{figure::prop_fee}.
    
Figure~\ref{figure::capacity} provides another interesting insight:
the capacities of the channels are surprisingly large:
around 5\% of the nodes invest a full Bitcoin into a channel.
One could interpret this result as quite a high level of commitment to the network. This also corresponds well with the
surprisingly low fees observed above. 

Figure~\ref{figure::degree} shows the number of channels per node.
We find that approximately 
5\% of the nodes are end-points (are of degree 1), and
25\% have degree 2.
The distribution is highly heavy-tailed and some nodes
exhibit very high degrees. Furthermore, about 88\% of the channels are connected to a node with degree higher than 600, and 90\% of the rest have the default configurations of base and proportional fee. This suggests that most of the channels that were created by the more sophisticated nodes, are almost always connected to the ``central'' nodes.

    \begin{figure} 
    	\centering
    	\includegraphics[scale=0.6]{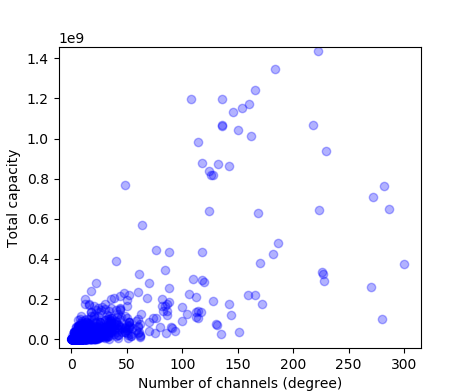}
    	\caption{Every point is a single node (5\% of the nodes was trimmed).}
    	\label{figure::degree_vs_capacity}
    \end{figure}

\subsubsection{Evaluation of Routing Properties}
    
\begin{figure} 
    	\centering
    	\includegraphics[scale=0.6]{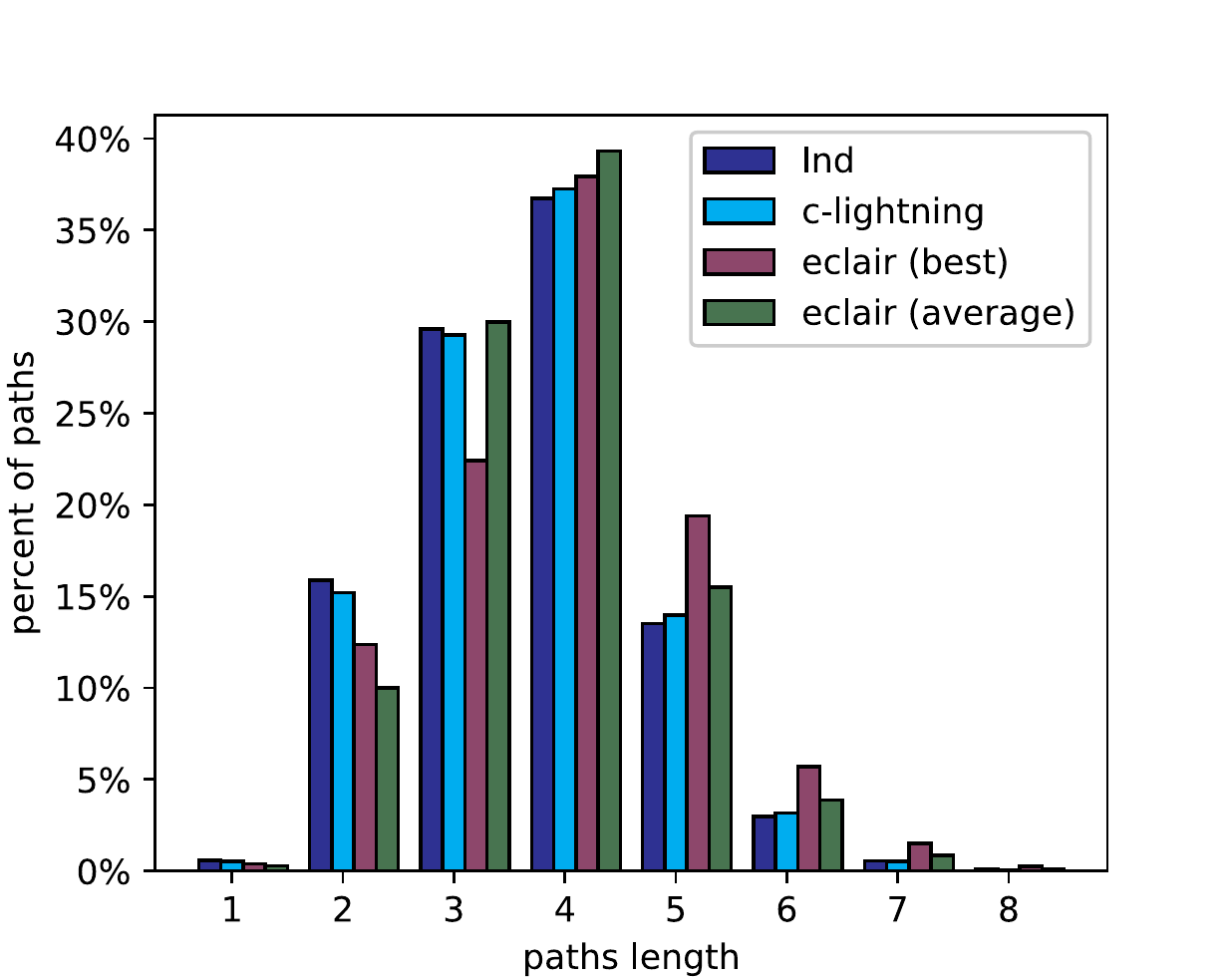}
    	\caption{Number of edges in each path for each implementation (for transactions of size 1000 millisatoshis)}
    	\label{figure::path_lengths}
    \end{figure}

We next take a deeper dive into routing properties. 
In particular, we examine the paths that are selected by the 
different routing algorithms used in the three main implementations. 
For every two nodes we determine the paths for 
transactions of size 1 satoshi (1000 millisatoshis).
Figure~\ref{figure::path_lengths} highlights the impact 
of the different routing algorithms nicely:
the path length distribution is affected by the specific algorithm. 

\begin{figure} 
	\centering
	\includegraphics[scale=0.6]{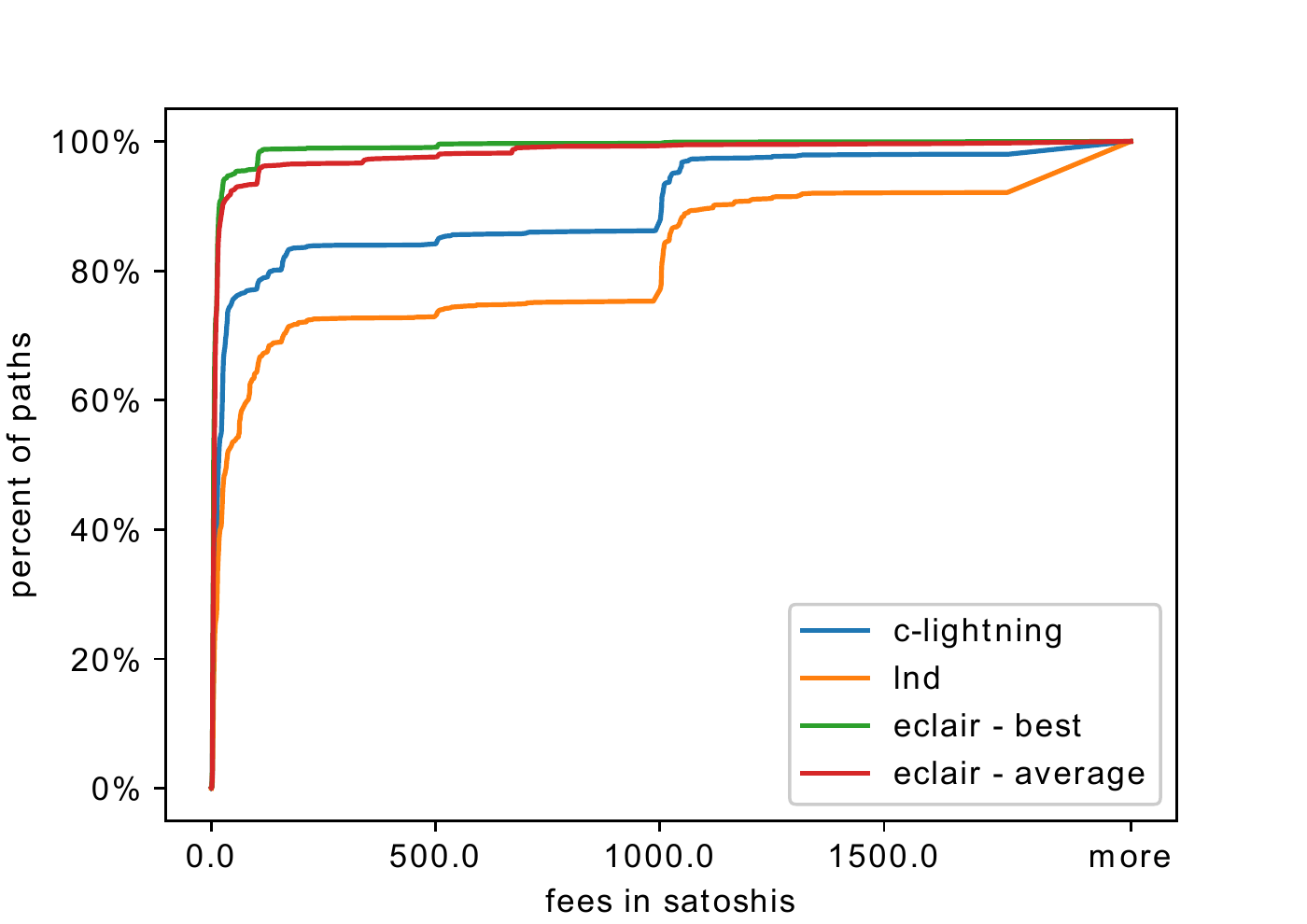}
	\caption{The fee volume for a transaction of size 1000 satoshis}
	\label{figure::fee_requested}
\end{figure}

Figure \ref{figure::fee_requested} shows the fee volume for the different implementations. We can see that Eclair is the ``cheapest'' implementation, and the reason is clear from the pseudo code; This is the only implementation that multiplies the channel's properties with the fee itself, where the others multiply by the total amount (transaction size + fee). 
    
\begin{figure} 
	\centering
	\includegraphics[scale=0.65]{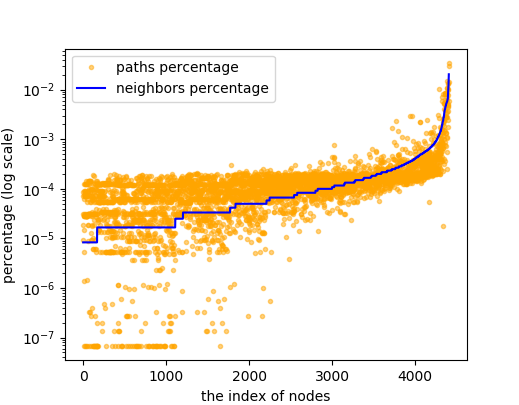}
	\caption{The correlation between the degree and the number of occurrences in paths. The nodes in this graph are ordered according to their degree.}
	\label{figure::degree_to_apparences}
\end{figure}

In Figure~\ref{figure::degree_to_apparences} we see the correlation between the degree of the nodes and the percentage of transactions that route through it. We see that in the lower degrees, there is a high variance in the percentage of paths. On the other hand, the variance decreases in the higher degrees, and the percentage of routes increases respectively.
    
\subsection{Feasibility of the Attack} \label{sec::hijack_analysis}

We now evaluate the feasibility of a DoS attack in two main scenarios:

\begin{itemize}
    \item \textbf{Collusion by existing central nodes:} We consider the case that a small number of highly central nodes collude and jointly launch a DoS attack using their existing resources and connections.  
    \item \textbf{External attacker:} An attacker joins the network, creates new channels to existing nodes and ``hijacks'' routing using low fees and other channel properties, competing with existing paths.
\end{itemize}

In what follows, let us assume that all pairs of nodes in the network connect 
to one another to transfer $1$ and $1000$ satoshis exactly once. For our analysis,
we count the number of disrupted pairs of transactions.

\subsubsection{Colluding Nodes}

\begin{figure} 
    	\centering
    	\includegraphics[scale=0.6]{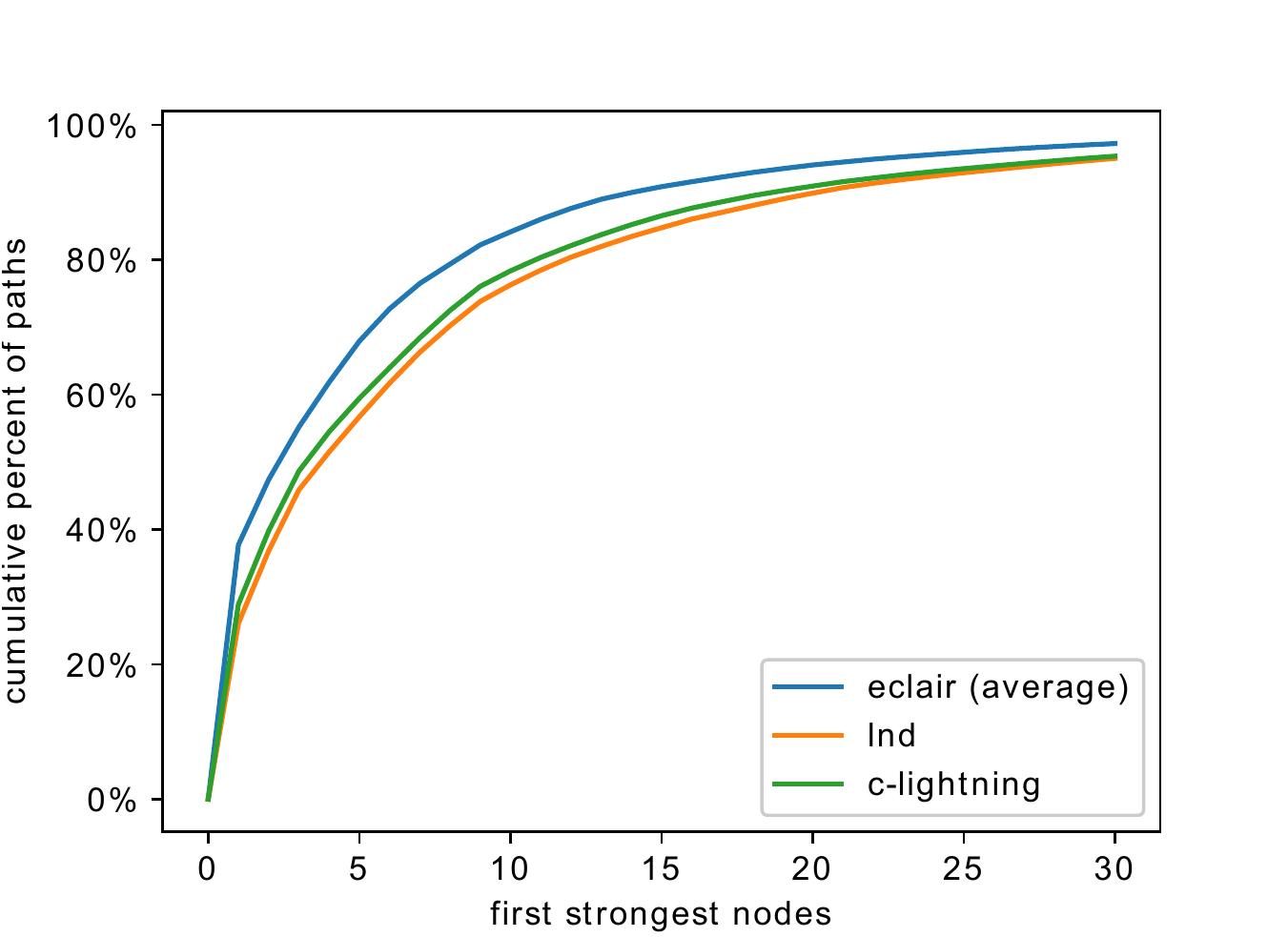}
    	\caption{percentage of paths that go through the most common nodes (assuming transaction sizes of 1000 millisatoshis).}
    	\label{figure::bribe_existing}
    \end{figure}

Figure~\ref{figure::bribe_existing} plots the nodes' centrality:
the number of paths going through the $k$ most central nodes (cumulative).
We can see that the five highest ranked nodes can disrupt roughly 60\% of all pair connections, and that differences between different implementations are relatively minor. 
Clearly, 
if these nodes collude and start a 
DoS attack, they will cause major disruptions to the network.
    
\begin{figure} 
    	\centering
    	\includegraphics[scale=0.63]{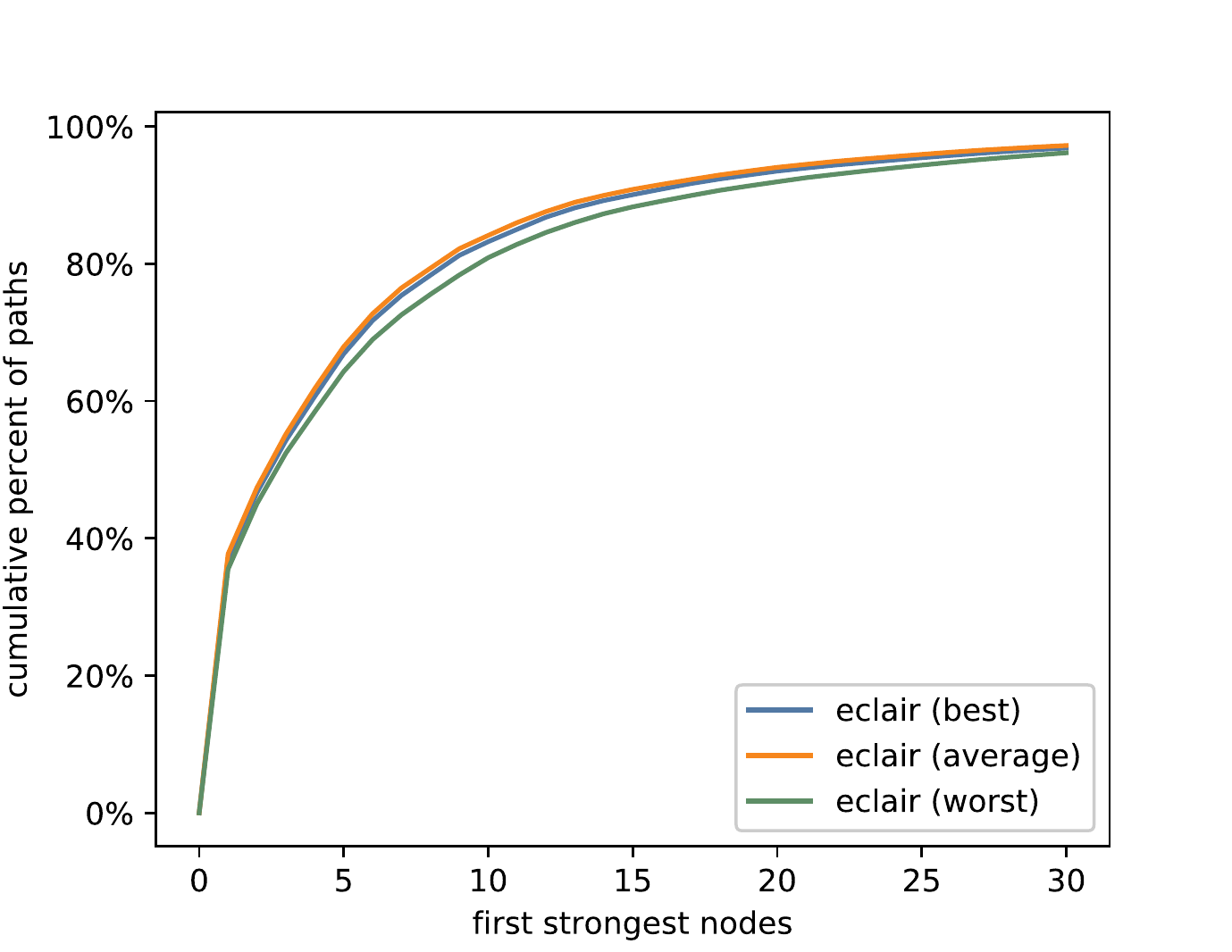}
    	\caption{Percentage of paths that go through the most common nodes. Average - increase the probability of hijack a created path, Worst - increase the probability to hijack nodes (every possible paths between the two).}
    	\label{figure::bribe_existing_eclair}
    \end{figure}

As Eclair's implementation chooses uniformly between the best three routes, we dive deeper with respect to that implementation. If some of the three top paths between a given pair of nodes do not pass through the attacker, there is a chance that a connection will form.
Therefore, in Figure~\ref{figure::bribe_existing_eclair},
we examine 3 metrics: (i) The fraction of hijacked best routes (lowest weight route of the 3 options), 
(ii) the fraction of pairs for which we hijack \textbf{\emph{all}} the top 3 routes, in order to build an attack that always works,
and (iii) the \emph{expected} fraction of hijacked routes from the top 3. 
The main lesson from the figure is that all metrics are very similar. Thus Eclair's randomization between the top 3 routes helps very little to avoid attackers. 

\begin{figure} 
    	\centering
    	\includegraphics[scale=0.28]{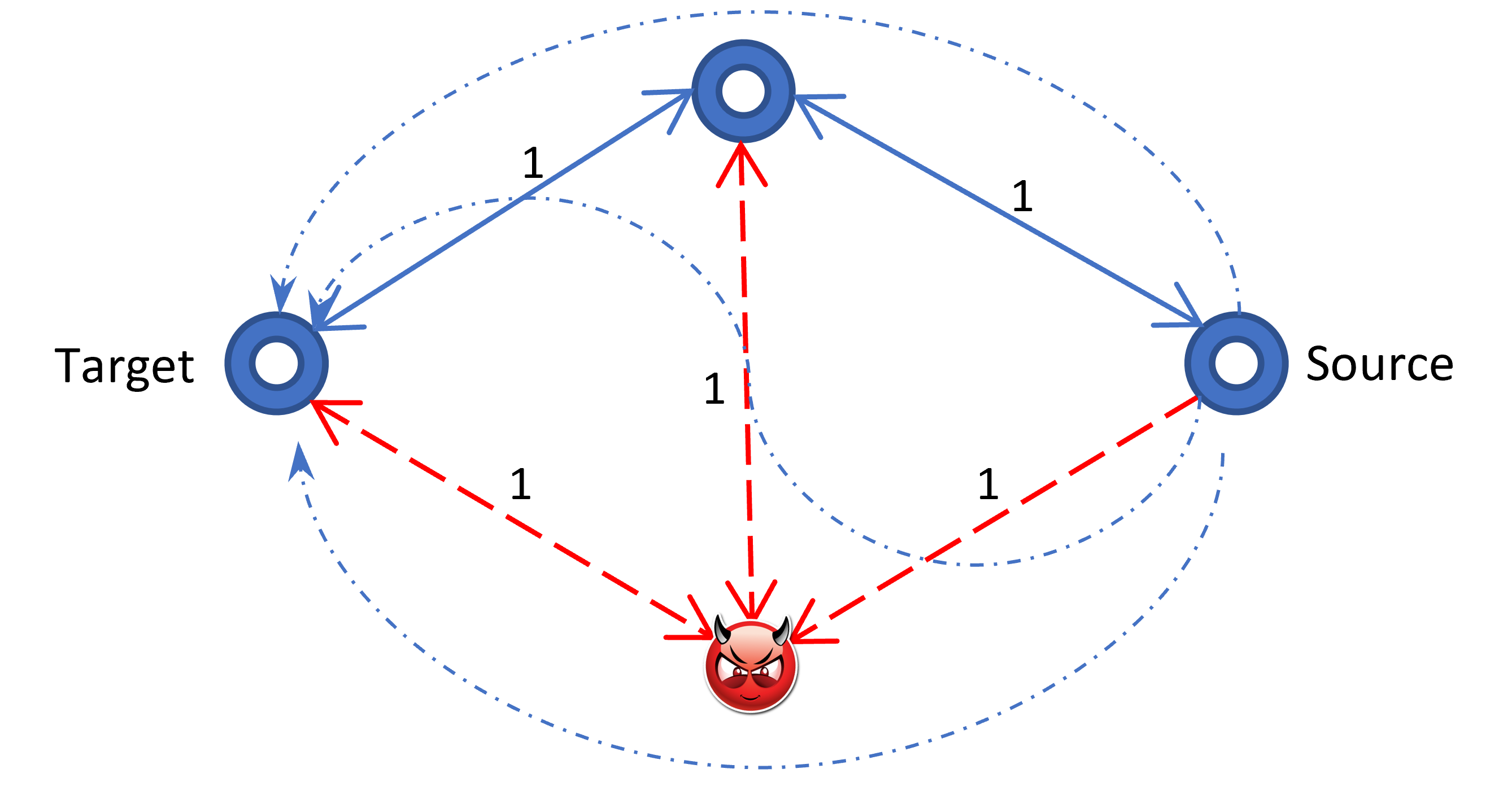}
    	\caption{If the attacker creates the red edges, the best 3 routes are illustrated in blue dashed lines. The approach to ``hijack the best path" results in the value $0.5$ (1 of 2 best paths pass through the attacker).
    	The approach ``hijack all the top 3" results in $0$ (as one path does not pass through the attacker, and ``hijack as many from the top 3" in $0.66$.}
    	\label{figure::explain_bribe_existing_eclair}
    \end{figure}

Digging deeper, the figure shows that 
(iii) yields the highest hijack rate, then (i), and the last is (ii). We try to illustrate an explanation in Figure~\ref{figure::explain_bribe_existing_eclair}.
    
    
\subsubsection{An External Attacker}

We now consider attacks by an external attacker that creates links to the network. 
Our discussion will focus on an attacker that tries to maximize the number of hijacked paths out of the the paths between all pairs of nodes, and not necessarily to maximize the fees it collects. 
    
\begin{figure} 
    	\centering
    	\includegraphics[scale=0.6]{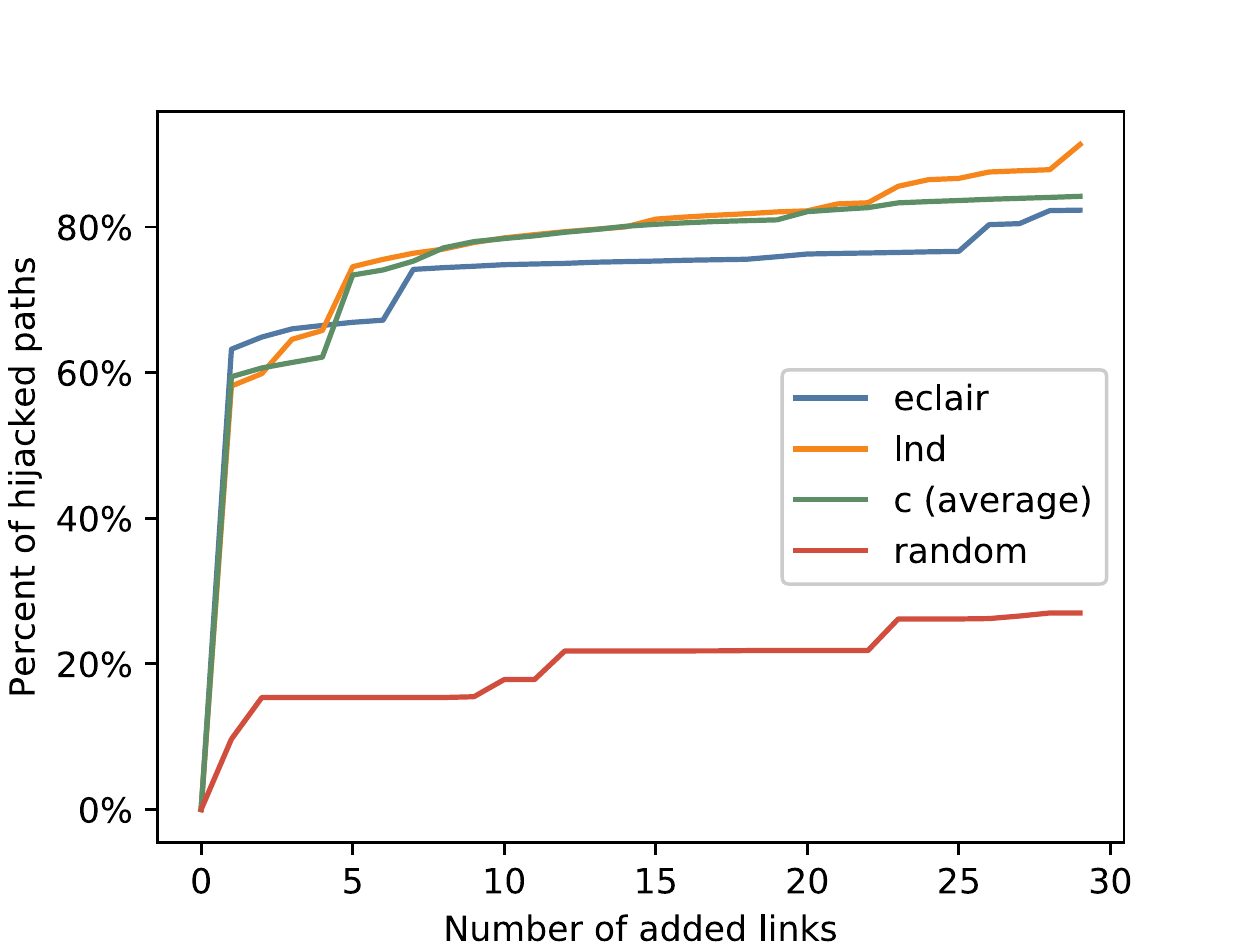}
    	\caption{Number of channels that we need to create (with zero fees and minimum delay) in order to hijack the paths of transaction sizes of 1000 millisatoshis.}
    	\label{figure::hijack_attack}
    \end{figure}

Figure~\ref{figure::hijack_attack} examines an attacker that creates new channels in order to attack the network. We used Algorithm~\ref{alg::implementation} that adds edges one by one, in a greedy approach, and calculated the percentage of the hijacked paths. We compared the attack impact between the different implementations. We added a control group, that was created by selecting connecting nodes uniformly and checking the hijack percentage in lnd. This graph is one of our main results. It shows the consequences of an attack on the network and compares between the different implementations.

\begin{figure} 
    	\centering
    	\includegraphics[scale=0.6]{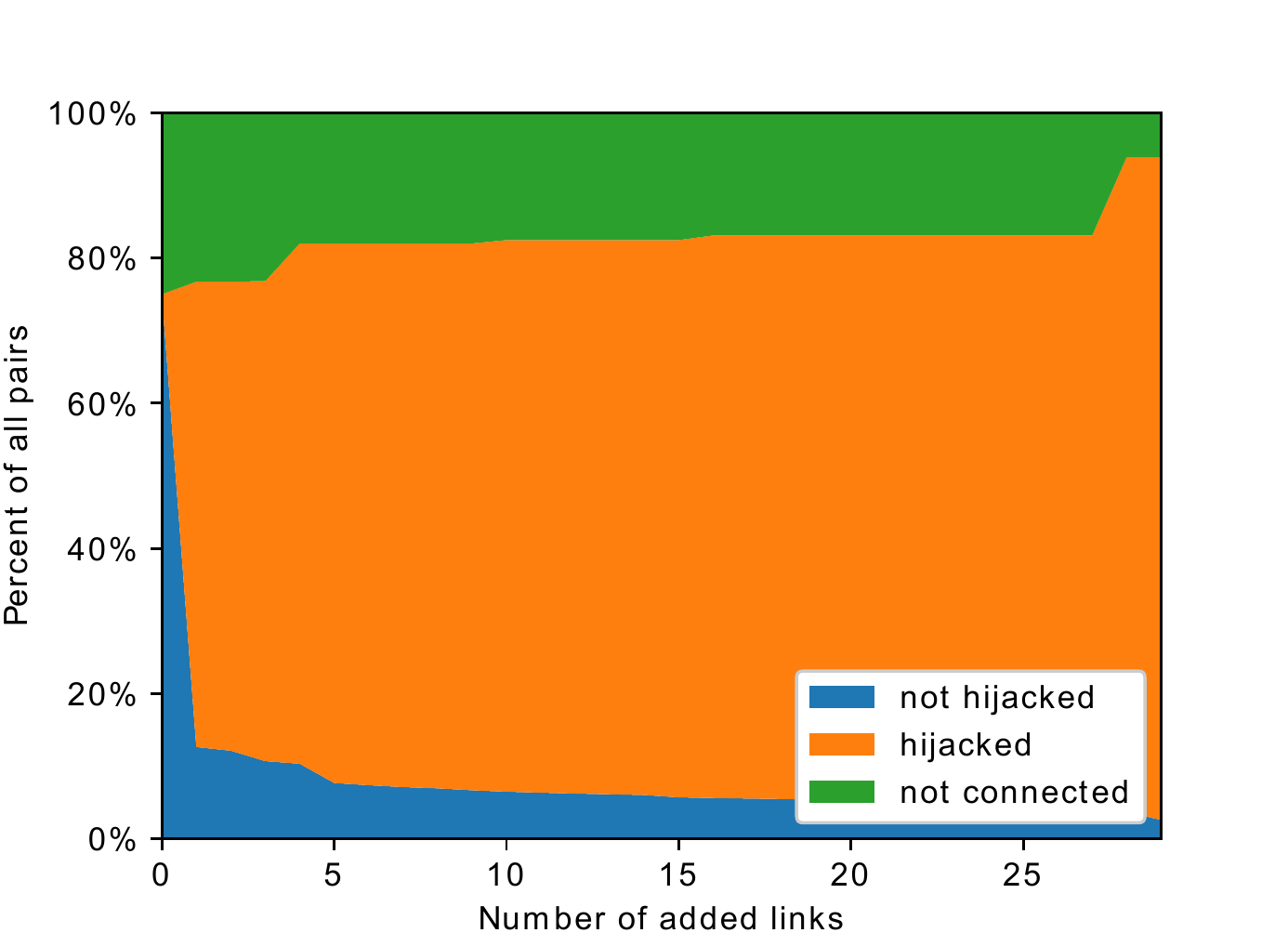}
    	\caption{Creating channels in the context of connecting new nodes (lnd).}
    	\label{figure::hijack_attack_new_paths}
    \end{figure}

Alternatively, we can consider the hijacked routes as part of the number of paths in the context of all the available connections between two nodes. Figure~\ref{figure::hijack_attack_new_paths} shows that the attack actually creates new available paths sometimes between nodes who were not previously connected. These new paths are now available 
as the route weight and hops are decreased (below the threshold of lnd).
    
\begin{figure} 
    	\centering
    	\includegraphics[scale=0.7]{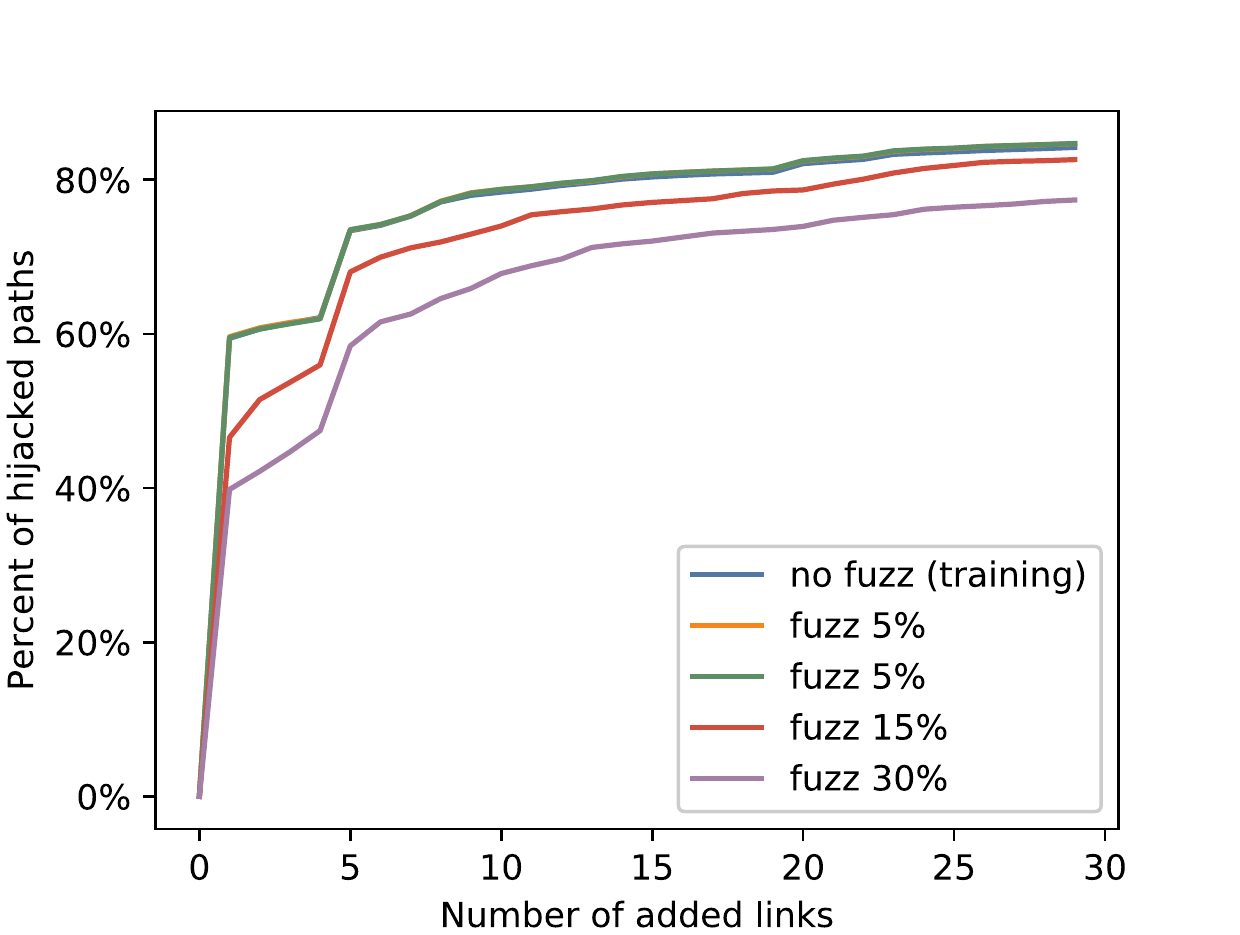}
    	\caption{The hijack percentage in the trained sampling (with no fuzz) 
    	compared to the percentage in other samples with fuzz = 5\% (default).}
    	\label{figure::hijack_attack_fuzz_impact}
    \end{figure}

Figure~\ref{figure::hijack_attack_fuzz_impact} shows the effectiveness of the weight-fuzzing method of C-lightning. We used the weight function without any fuzzing in order to greedily find the channels that the attacker should create, and then we evaluated our results against routing with the default fuzz parameter. We repeated this 4 times with different fuzz rates. The figure indicates that the re-introduction of the default fuzz ($\pm 5 \%$) does not prevent the attack. Our suggested explanation is that the fuzz multiplies
only the channel's fee, which is very low in this attack, and thus does not substantially change routing decisions. 
    
\subsection{Amplified Attack With Delays}

This subsection discusses a way to amplify the DoS attack, by increasing the time that the attacker holds the hijacked transaction (the delay parameter). Here, we suggest the following enhancement: the attacker will report a high delay value for his node, which will then affect the delay of all preceeding HTLCs in the path (recall that delays must accumulate in the reverse order of the path to guarantee intermediary nodes that the outgoing HTLCs expire before the incoming ones).

\begin{figure} 
	\centering
	\includegraphics[scale=0.6]{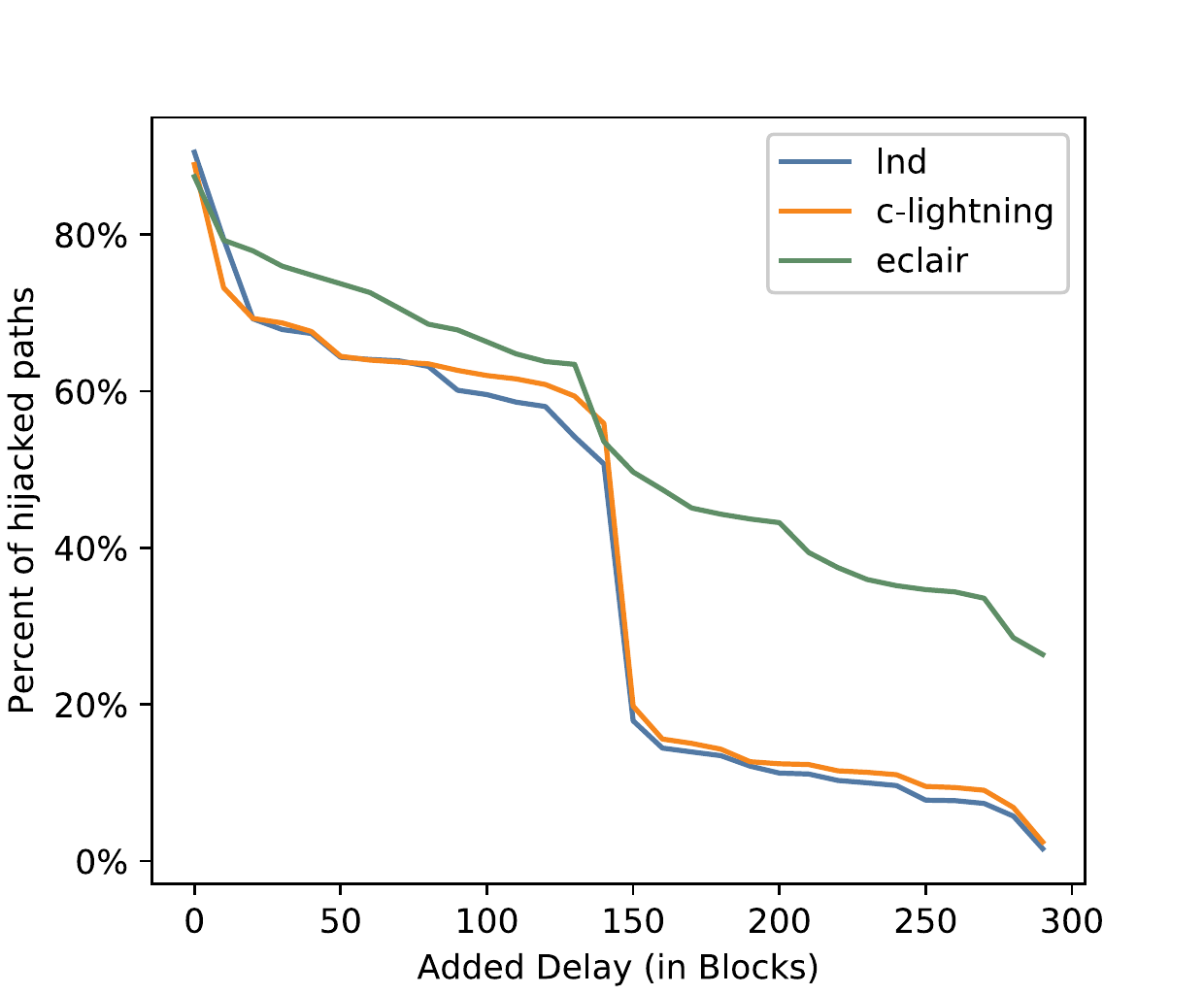}
	\caption{The hijack percentage when we create edges for increasing delays, using the top 30 new links from Figure~\ref{figure::hijack_attack_new_paths}}
	\label{figure::delay_attack}
\end{figure}

Note that there is a trade-off for the attacker, because this delay is part of the channel's properties that are used in order to calculate the weight, so higher delay means a stronger effect on fewer nodes. Figure~\ref{figure::delay_attack} shows the hijack rate when the attacker increases the delay of the 30 channels that were created in the previous subsection. Note that there is a big drop around the delay of 144 blocks, which correlates with the fact that many nodes use this as their parameter (Figure~\ref{figure::delay}). 

\section{Analysis and Optimization\\of Attacker Strategy}\label{sec:preliminaries}

Having demonstrated the feasibility of the attacks empirically,
we now explore the optimization problems  underlying the attack
from an algorithmic perspective. To this end, we propose
an analytical model for the adversary.
In particular, we will show that while determining the best adversarial
strategy is an NP-hard problem, efficient polynomial-time approximation
algorithms exist. To this end, we establish an interesting connection
to centrality theory, which turns out to come with a twist in our setting.

\subsection{Preliminaries}

Let $V$ be the nodes that participate in the network and $E$ be the channels, 
i.e., $(u,v) \in E \subseteq V \times V$ are nodes with established channel\footnote{Note that we 
are not interested in the P2P network itself, but only the channels graph.}. 
A valid path from a source node $s \in V$ to a target node $t \in V$ is a list of edges $((u_1, v_1), \cdots, (u_n, v_n)) \in E^*$ where $u_1 = s, v_n = t$ and $v_i = u_{i+1}$ for all $i$. 

Path selection algorithm $\alg_W$ is an algorithm with the inputs: source node, target node, and the channels' graph.
It returns a valid path from the source to the target. 

The centrality of a channel $e$ is the percentage of the network's transactions 
that pass through this channel. In the same way, define the centrality of a set of channels $e_1, \cdots, e_n$ (note: this is not necessarily the sum). Denote this function with $\cent: 2^E \longrightarrow \mathbb{R}$. 
Note that although this is a close notion to the \emph{betweeness centrality}
in the literature, we consider here routing algorithms that do not necessarily choose the shortest path (lowest weight), like Eclair's top-k randomization. 

\subsection{Attacker's Algorithms}
    
In general, computing an optimal attack is hard,
as the problem of computing optimal link
additions is already NP-hard for shortest paths,
i.e., betweenness centrality~\cite{bergamini2018improving}.
We hence explore the possibility of polynomial-time
approximation algorithms: algorithms which are fast
enough to scale at least to 
all the nodes and channels of the Lightning network (about 4000 nodes at the moment).
   
  In the following, we will explore the opportunity introduced
  by submodularity, and consider the connection to 
  the problem of betweeness maximization with bounded budget~\cite{bei2011bounded}.
    
    \begin{lemma}\label{lemma:cent_submodular_existing}
        The centrality rate of \textbf{existing edges} for a given node is a 
        non-negative, monotone, sub-modular function. That is, 
        for $\forall A,B \subseteq E$ it holds that $\cent(A) + \cent(B) \ge \cent(A \cup B) + \cent(A \cap B)$
    \end{lemma}
    \begin{IEEEproof}
        Recall that $\cent(A)$ is the number of transactions that 
        go through the channels in A. The non-negative and monotone 
        properties follow directly from the definition. 
        Regarding the sub-modularity, we consider the
        different cases:  
        (i) Transactions that go through only 
        one of $A$ and $B$, we count exactly once 
        on both sides of the equation. (ii) Transactions that 
        do not go through $A$ or $B$, we do not count on both sides.
        (iii) Transactions that go both through $A$ and $B$ and that are in $A \cap B$,
        we count twice on every side. (iv) Transactions that go both through $A$ and $B$,
        but that are not in $A \cap B$, we count twice on the left side, but only once on
        the right.
        Overall, we obtain that the left side can be larger then the right side, 
        as desired.
    \end{IEEEproof}
    \begin{remark}
        The above lemma can be rephrased also to $A,B \subseteq V$ (group of nodes instead of edges).
    \end{remark}
    
    \begin{lemma}\label{lemma:cent_submodular_new}
        The centrality rate of creating \textbf{new edges} for 
        a given node is a non-negative, monotone, sub-modular function. 
        I.e.~$\forall N_1 \subseteq N_2 \subseteq V$ and $x \in V - N_2$,  denote by $A,B$ the groups of new channels that connect
        $N_1,N_2$ to a new node $v$, respectively, and $e$ that connects $x$ to $v$. 
        Then it holds that $\cent(A \cup \{e\}) - \cent(A) \ge \cent(B \cup \{e\}) - \cent(B)$.
    \end{lemma}
    The proof is equivalent to the proof of Theorem~5.2 in \cite{bergamini2018improving}.
    The key ideas are: (i) If we consider two sets of new edges $X \subseteq Y$, then the distance between every two nodes in the graph with the new edges from $X$ is greater equal 
    the distance when adding $Y$. (ii) If all the new edges are connected to only 
    the attacker's nodes, then the attacker's centrality when adding $X$ is 
    lower than when adding $Y$. (iii) Strong inequality in (i) 
    implies strong inequality in (ii).
    
    Let us now consider a
    repetitive attack, in which our goal is to attract others 
    to always route through our node. To achieve this goal, we will 
    add many edges with 0 fees and delay. Each such channel bears some costs for the attacker due to the need to lock funds by the attacker. To decrease costs we therefore wish to minimize the number of channels. 
    
    \begin{algorithm}
        \DontPrintSemicolon
        \SetKwProg{Fn}{Function}{:}{}
        
        \SetKwFunction{FGreedy}{GreedyApproach}
        \Fn{\FGreedy{$k$, $f$, $E, V$, $\bar{v}$}}{
             \For{$i=1, \cdots, k$}{
              $e = \arg\max_{v \in V} f(E \cup \{(v, \bar{v})\})$ \;
              $E = E \cup \{e\}$ \;
             }
         }
         \caption{Greedy perspective to find $k$ channels that maximize the function $f$}
         \label{alg::general_greedy}
    \end{algorithm}

    \begin{theorem}
        A greedy algorithm that, given edges $E$, node $n$ and number $k$, 
        iteratively finds the edge $e$ that maximizes the centrality rate of $n$ and updates 
        $E = E \cup \{e\}$ (Algorithm~\ref{alg::general_greedy}), 
        gives a $1-(1-\frac{1}{k})^k$ approximation.
    \end{theorem}
    \begin{IEEEproof}
        As in Section 4 (corollary of Prop. 4.3) of \cite{nemhauser1978analysis}, 
        we implied the greedy heuristic on the function $\cent$ (which is a sub-modular 
        set function according to Lemmas~\ref{lemma:cent_submodular_existing},\ref{lemma:cent_submodular_new}).
    \end{IEEEproof}
    
    It remains to show an efficient method to calculate
    $\arg\max_e f(E \cup \{e\})$. 
    This can simply be achieved by dynamic programming: 
    find the best edge to add, and update the state accordingly.
    Algorithm~\ref{alg::implementation_naive} describes this idea.
    
    In our algorithm, we made some further improvements, 
    based on the fact that there are no
    valid paths between \textbf{all} the pairs (because of defaults of max hops or max fee). 
    See Algorithm~\ref{alg::implementation} for details.
    
    \begin{algorithm}
        \DontPrintSemicolon
        \SetKwProg{Fn}{Function}{:}{}
        
        \SetKwFunction{FPreProc}{Preprocessing}
        \Fn{\FPreProc{$E$, $V$}}{
             $dbPaths = \emptyset$ \;
             $dbVertexes = \emptyset$ \;
             \For{$v \in V$}{
                $dbPaths$.update(perform dijkstra and get shortest paths and weights to $v$) 
             }
             $dbVertexes$.update(map between vertex to all the participated paths)
        }\;
         
        \SetKwFunction{FFindNextNaive}{FindNextNaive}
         \Fn{\FFindNextNaive{$E$, $V$, $\bar{v}$}}{
            $best$, $value$ = null, 0; \;
            \For{$candidate \in V$}{
                $counter$ = 0 \;
                \For{$src, dst \in V \times V$}{
                    \If{shortest(src, $candidate$) + shortest($\bar{v}$, dst) $\le$ shortest(src, dst)} {
                        $counter$ ++ \;
                    }
                 }
                 \If{counter $>$ value} {
                    $best$, $value$ = $candidate$, $counter$ \;
                 }
             }
             \Return $best$ \;
         }\;
         
         \caption{calculate $\arg\max_e f(E \cup \{e\})$ efficiently}
         \label{alg::implementation_naive}
    \end{algorithm}
    
        \begin{algorithm}
        \DontPrintSemicolon
        \SetKwProg{Fn}{Function}{:}{}
        
        \SetKwFunction{FFindNext}{FindNext}
         \Fn{\FFindNext{$E$, $V$, $\bar{v}$}}{
            $best$, $value$ = null, 0; \;
            \For{$candidate \in V$}{
                $counter$ = 0 \;
                \For{$src$ reachable to $candidate$}{
                    \For{$dst$ reachable from $src$}{
                        \If{($src,dst$) is already attacked}{continue}
                        \If{shortest(src, $candidate$) + shortest($\bar{v}$, dst) $\le$ shortest(src, dst)} {
                            $counter$ ++ \;
                        }
                    }
                 }
                 \If{counter $>$ value} {
                    $best$, $value$ = $candidate$, $counter$ \;
                 }
             }
             \Return $best$ \;
         }\;
         
         \caption{Our implementation of findNext, while optimizing the runtime and reduce calls to the db}
         \label{alg::implementation}
    \end{algorithm}

    It is important to notice that the above algorithms are indeed not optimal and are just an approximation. This only strengthens our results: these algorithms yield, in practice, very good results (for the attacker), and more sophisticated attackers may do even more damage.
    
\section{Exploring Solutions} \label{sec::suggested_solution}

Let us now explore methods that can increase the robustness of the network,
and at least partially address the tradeoffs observed above.
In the following, we will suggest two different solutions:
the first is based on a game-theoretic perspective where 
we analyze the strategies of the attacker and defender as two rational players.
The second is based on a set of conclusions that we learned from the above experiments.

\subsection{Game Theoretic Approach}

We can reason about the interaction of the attacker and the defender 
as a continuous game, where the attacker tries to sabotage 
as many edges as possible in a long time range. On the other hand, 
the attacker may try to perform ``fast attacks", where the adversary gains access to 
much resources for a short period, and then
tries to block the transactions in the network. As for the defender's perspective,
the matching examples are if 
it performs cycles of trust-building or maximizes the security of each transaction.

Another interesting approach to see the interaction is to consider 
non-selfish nodes, where we utilize the interaction between a single attacker 
and the network as a whole. The nodes cooperate in order to increase the overall security of the network. This approach is interesting because it must include other incentives to the nodes
(otherwise they will employ selfish routing).

In the following, as a first step, 
we will examine only a very simple approach, 
in which the defender is selfish, and the attacker ties to attack a single transaction.

\boldheader{Simple example}
We will present here a short game theoretic analysis of a specific case, 
wherein a selfish node (defender) tries to perform a single transaction. 
On the other hand, the attacker is trying to attack this node specifically, 
knowing the exact target node of the transaction and the transaction size. 
We assume that they both are fully rational.

The model that we suggest here is simple: assume that the
weight function is simply the sum of fees on channels on the route, 
and that the attacker can always create a channel that has less weight 
compared to an existing channel. For the attacker, the price to create nodes 
is negligible, and the price to establish a channel with capacity of $c$ is $c \cdot I$, 
where $I$ is a global constant that represents liquidity costs (i.e., 
the attacker pays the interest rate on locked funds that are otherwise unused 
for the duration of the attack)\footnote{Note that we omit the time span for which 
the money is locked. This obviously may change $I$. We will get back to this point in the next subsections.}. Moreover, assume that in the case of successful attack, the attacker wins the same value that the defender looses. Denote this value by $H$. The defender, tries to minimize the weight to execute the transaction.

The pure strategies available to the defender are described by all the paths to the target, 
and the attacker's pure strategies are always to add new nodes and new channels.
The utility function is thus defined as follows: if the attacker created the channels $C_{att} = a_1, \cdots, a_k$ and the defender chooses the path $C_{def} = d_1, \cdots, d_l$, then the utility of the attacker is $U_{attacker} = H \cdot \delta_{C_{att}, C_{def}} - I \cdot \sum_{c \in C_{att}} c_{capacity}$, and the defender's utility is $U_{defender} = -H \cdot \delta_{C_{att}, C_{def}} - \sum_{c \in C_{def}} c_{fee}$, where $c_{capacity}, c_{fee}$ are the capacity and fee of the channel $c$, and $\delta_{A,B}$ is 0 if $A \cap B = \emptyset$ or 1 otherwise.

Intuitively, in every Nash equilibrium in this game, 
the attacker will hijack the paths that the defender will use with the highest probability. On the other hand, the defender will try to minimize the probability to use every set of specific channels (not being predictable).

Exploring examples in different graphs is interesting,
but beyond the scope of this paper and left for future research. 

\subsection{Lesson Learned - Suggested Weight Function}
We next suggest first ideas 
based on the empirical experiments that we did in Section~\ref{sec::hijack_analysis}. 
We will focus on insights that aim to increase the cost of a successful hijack attack, 
and we hope that these insights will be considered when implementing a new weight function. 
We later give an example of the impact of slight changes to Eclair's weight function.

The first lesson is related to the vulnerability of Eclair to the delay attack. 
Here, the weight is determined by multiplying the channel's parameters with the fee. 
This creates a tradeoff between the fee and the delay: when multiplying
the delay and dividing the fee by the same factor, this will result in the same weight, 
although the more intensive attack. Therefore we suggest to either multiply the delay by the total amount of the transaction or to summarize it to the other evaluations.

The second lesson is how to create a non-deterministic algorithm. 
C-lightning adds noise to the channel's fee (fuzzing). As we saw, it has a
low effect in case the attack because of the exceptionally low fees of the attack channels. On the other hand, Eclair chooses uniformly a path among the top ones (and not necessarily the best), and that has also a low effect in a case of an attack because of the amount of different paths that the attacker 
can create with a small amount of resources. Therefore we suggest to add fuzz to the total weight of the channel, and avoid choosing one of the top-$k$. The difference between these two options is rather 
that the attacker needs to be better than the others by a small constant difference or linearly.

Few more lessons worth pointing: (i) older nodes are better because the interest rate (as a parameter in the utility of the attacker) is higher; (ii) high capacity is safer than low capacity; (iii) the delay is important.

We were considering to add the betweeness rank to the weight calculation, 
but we think that it makes the routing algorithm computationally expensive, and therefore diverges 
from the goal to create simple improvements to existing weight functions.

\boldheader{Improve Eclair}
We think that creating a weight function that will be resilient to hijack attacks and preserve important network properties (such as connectivity, low fees, etc.) should be researched properly. In the next discussion we will not suggest an optimal weight function, but only try to improve the existing weight functions using the same structure of implementation. We based our suggestion on Eclair's weight function that was presented in Section~\ref{subsubsec::imp_eclair}.

We suggesting the improvement of taking into account the following channels properties. 
The weight parameters should be determined according to the network and the user's configuration. The general structure is:
$$ scale = \mathcal{N}(1, \sigma^2) $$
$$ fee = ams[i+1] \cdot p[i].propFee + p[i].baseFee $$
\begin{align*}
 we&ight[i] = scale \cdot (\\
 &normalizedDelay \cdot delayRatio \\
 &+ normalizedHeight \cdot ageRatio \\
 &- normalizedCapacity \cdot capacityRatio \\
 &- capacity \cdot height \cdot IntrestRatio \\
 &+ \frac{fee}{ams[i+1]} \cdot feeRatio \\
 )
\end{align*}
For $\mathcal{N}(1, \sigma)$ which is the standard Gaussian distribution around $1$ 
with variance $\sigma^2$ and some normalization factors and ratios. Note that the fee is not normalized, the scale multiplies everything and the negative sign in the capacity parameters.

\begin{figure} 
	\centering
	\includegraphics[scale=0.55]{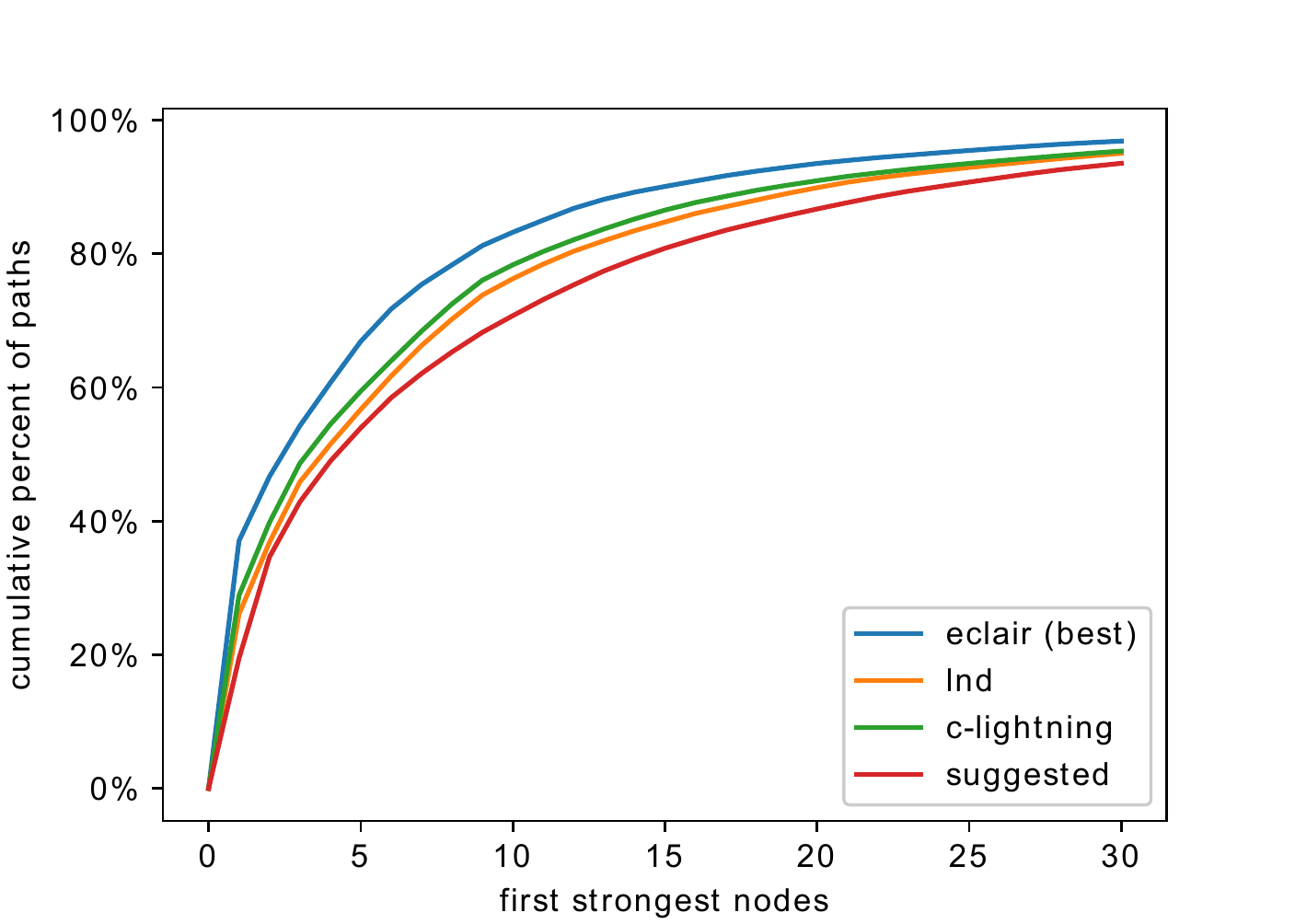}
	\caption{Percentage of paths that go through the most common nodes (assuming transaction sizes of 1000 millisatoshis).}
	\label{figure::suggested_bribe_attack}
\end{figure}

\begin{figure} 
	\centering
	\includegraphics[scale=0.6]{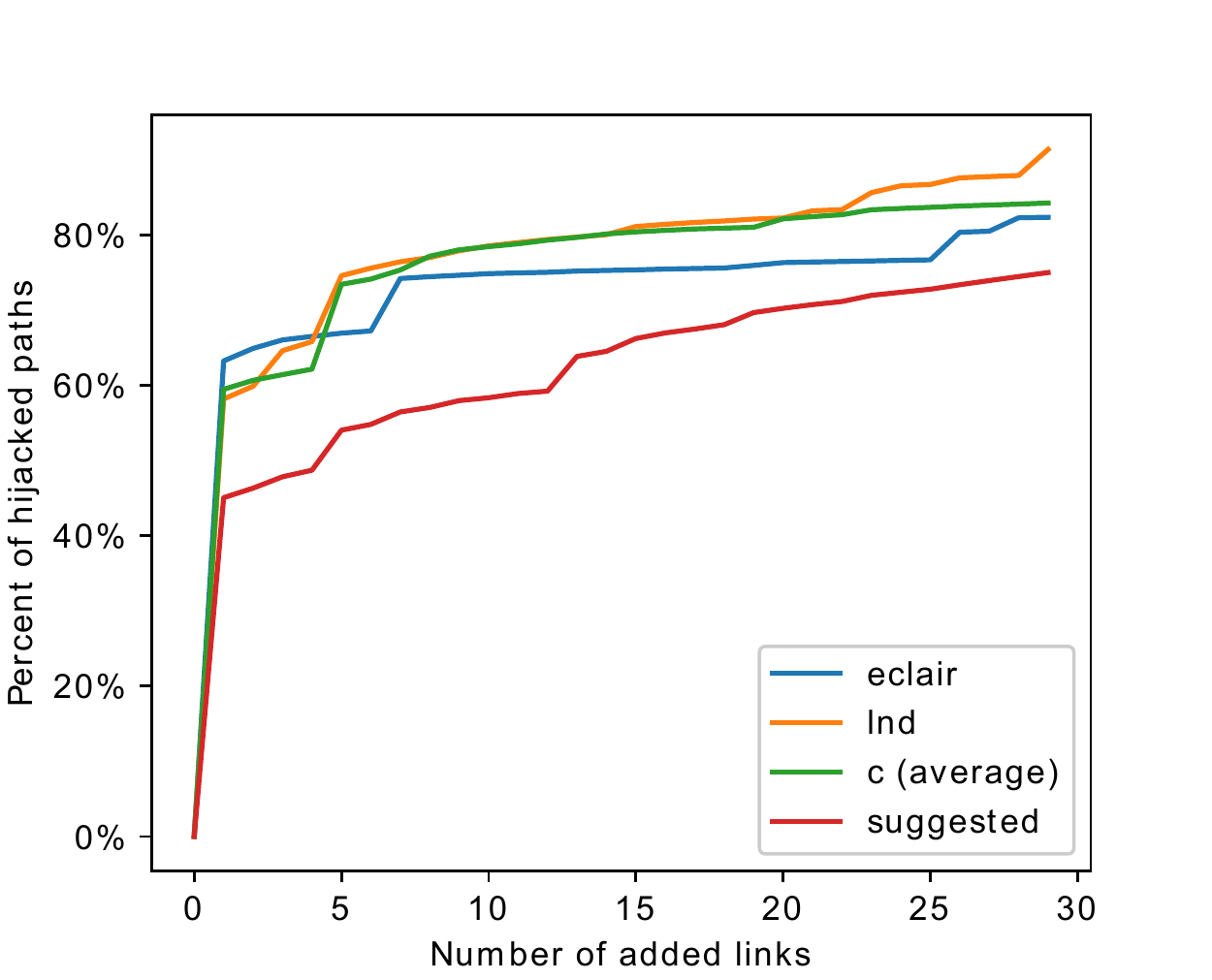}
	\caption{The hijack percentage in the trained sampling (with no fuzz) comparing to the percentage in other samples with fuzz = 5\% (default).}
	\label{figure::suggested_add_links_attack}
\end{figure}

\begin{figure} 
	\centering
	\includegraphics[scale=0.6]{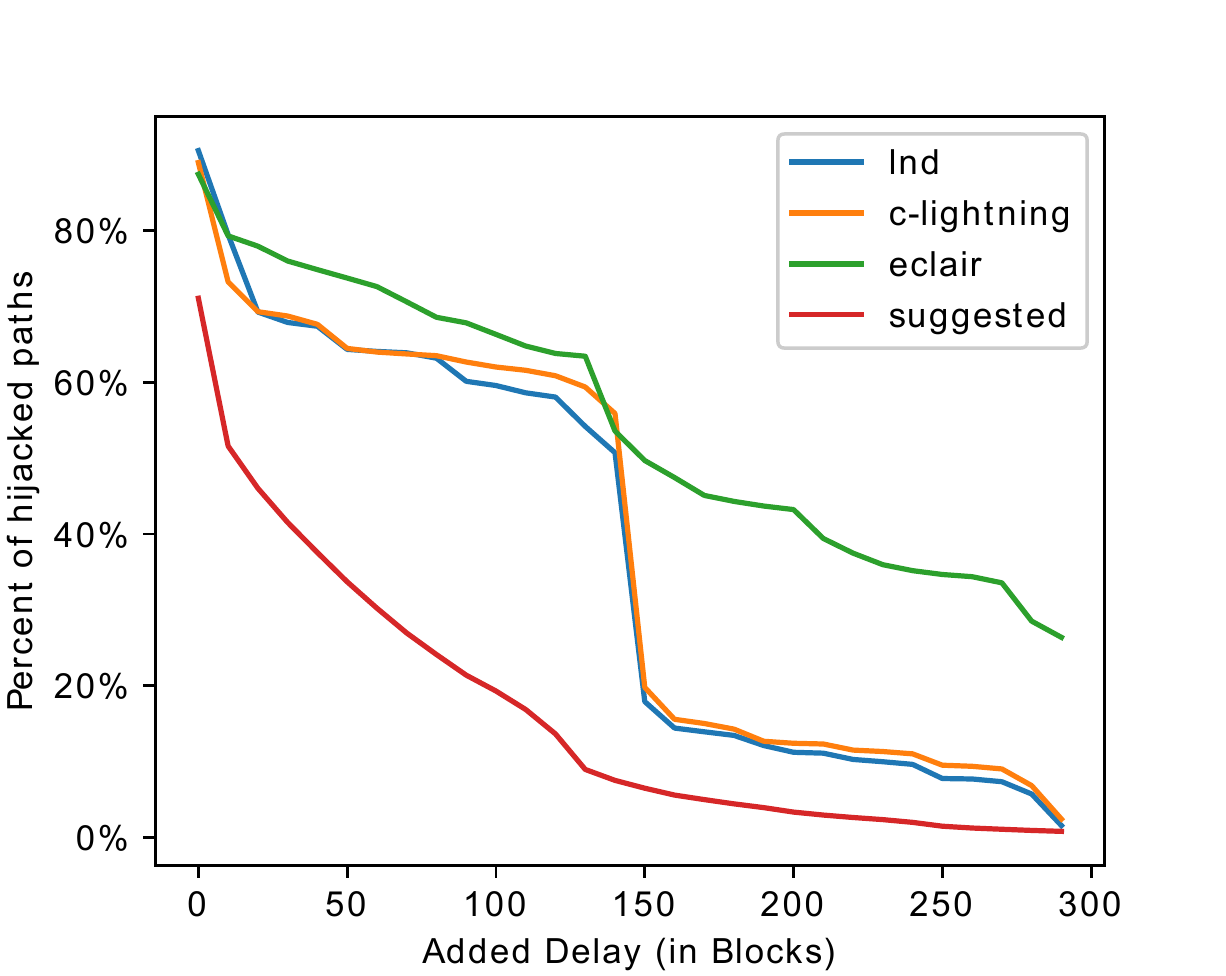}
	\caption{The hijack percentage when we create edges with higher and higher delay, using the top 30 new links}
	\label{figure::suggested_delay_attack}
\end{figure}

In order to evaluate 
the above function, we took the following parameters: $\sigma = 0.2, delayRatio=0.5, ageRatio=0.5, capacityRatio=0.3, feeRatio=100$, and the normalization parameters of Eclair.
We implemented this weight function and evaluated it using the same experiments as before. The results are presented in Figures~\ref{figure::suggested_bribe_attack},\ref{figure::suggested_add_links_attack},\ref{figure::suggested_delay_attack}. 

We note that these results require further exploration, specifically, it is important to evaluate other features of the new weight function, including the average fees for paths that it finds, and the failure rates of paths it selects due to liquidity imbalances. We leave such deeper evaluations for future work. 

% OLD FUNCTION
% $$ scale = 1 + fuzz \cdot (2 \cdot \frac{h}{2^{64}-1} - 1) $$
% $$ fee = ams[i+1] \cdot p[i].propFee + p[i].baseFee $$
% \begin{align*}
%  weight[i] = (ams&[i+1] + fee) \cdot scale \cdot\\
%  (&normalizedDelay \cdot delayRatio \\
%  &+ normalizedHeight \cdot ageRatio \\
%  &+ normalizedDegree \cdot degreeRatio)
% \end{align*}
% For upper and lower bounds: 
% \begin{align*}
%     9 &< delay < \num{2016} \quad delayRatio = 0.2 \\
%     0 &< height < \num{8640} \quad ageRatio = 0.5 \\
%     30 &< degree < 600 \quad degreeRatio = 0.3
% \end{align*}


\section{Related Work}\label{sec:relwork}

Since Bitcoin was first deployed in 2009 \cite{nakamoto2008bitcoin}, 
it received significant interest in academia, including security aspects.
While the main initial security research focus was on 
the analysis of the double-spending attack \cite{rosenfeld2014analysis},
many additional vulnerabilities were identified later~\cite{li2017survey}.

Also the P2P network has been analyzed intensively, 
for Bitcoin \cite{koshy2014analysis} and Ethereum \cite{kim2018measuring} (including the
centralization analysis in \cite{gencer2018decentralization}), 
but similar analysis existed already, e.g., 
for Skype \cite{guha2005experimental}. Our routing attacks can generally
be understood from the perspective of centrality; as discussed,
especially betweenness centrality has been studied much in the literature, as well as its interesting generalizations~\cite{dolev2010routing}. However, our model is different because the weight of each edge is determined by all the path (and not only by the edge). More implicitly, the weight of the edge is calculated using the amount of millisatoshis that it is transfer, and this amount could be changed if we will choose a different path (because of the difference in fees).

Attacks on the network level are also known, e.g.,
the eclipse attacks on Bitcoin \cite{heilman2015eclipse} and Ethereum \cite{marcus2018low}, 
or the attacks possibly performed by ASes \cite{apostolaki2016hijacking}. 
Route hijacking attacks were researched in a variety of fields, such as wireless ad-hoc networks \cite{deng2002routing}, general P2P networks \cite{naoumov2006exploiting}, and Bitcoin~\cite{apostolaki2017hijacking}.

An interesting recent work also discusses 
path hijacking in the Lightning network~\cite{rohrer2019discharged}.
There, the focus is on isolation attacks: 
the authors consider only the graph of the channels, without referring to the different implementations of the routing algorithms. Our work continues this idea and generalizes it to a general DoS
attack, where the attacker tries to damage the transactions of the network and not the nodes themselves. Therefore the delay amplifier and the analysis of the differences between the weight functions and 
randomized path selection has not been researched yet.

\section{Conclusion}\label{sec:conclusion}

This paper identified a novel attack on off-chain networks 
which introduces an interesting tradeoff both for the attacker 
as well as the rational defender. We have demonstrated the feasibility of this
attack on different networks and provided a first analysis.

We showed an empirical difference between the existing methods used
to randomize the chosen path. Random fuzzing on the fee of every channel was found to yield, in practice, weak protection against this type of attack compared to fuzzing the overall weight of the channel. Furthermore, we showed that if the defender considers the fee as a multiplier to the weight, then it will be especially vulnerable to the increasing delay attack.
We also proposed a first game theoretic model
for designing a new weight function.
%such as the capacity and age (which are the main costs of the attack).

We see our work as a first step and believe that it opens several interesting
avenues for future work. Generally, it will be interesting to analyze properties that weight functions should have, and build optimal functions accordingly.
It would also be interesting to consider the use of mechanism design
to incentivize nodes to choose routes that will increase the overall security of the network. 
Finally, it will be interesting to examine this attack on future features,
%that aim to decrease the knowledge of each node on the rest of the network. In particular, 
such as node ``switchboards" for message passing: the attacker may connect only to them to  control all the nodes that use it.

%\todo{Note: NDSS page limit is 13 + references}

\bibliographystyle{plain}
\bibliography{main}

\begin{thebibliography}{10}

\bibitem{BOLT}
Basis of lightning technology ({BOLT}s).
\newblock https://github.com/lightningnetwork/lightning-rfc.

\bibitem{apostolaki2016hijacking}
Maria Apostolaki, Aviv Zohar, and Laurent Vanbever.
\newblock Hijacking bitcoin: Large-scale network attacks on cryptocurrencies.
\newblock {\em arXiv preprint arXiv:1605.07524}, 2016.

\bibitem{apostolaki2017hijacking}
Maria Apostolaki, Aviv Zohar, and Laurent Vanbever.
\newblock Hijacking bitcoin: Routing attacks on cryptocurrencies.
\newblock In {\em 2017 IEEE Symposium on Security and Privacy (SP)}, pages
  375--392. IEEE, 2017.

\bibitem{armknecht2015ripple}
Frederik Armknecht, Ghassan~O Karame, Avikarsha Mandal, Franck Youssef, and
  Erik Zenner.
\newblock Ripple: Overview and outlook.
\newblock In {\em International Conference on Trust and Trustworthy Computing},
  pages 163--180. Springer, 2015.

\bibitem{bamert2013have}
Tobias Bamert, Christian Decker, Lennart Elsen, Roger Wattenhofer, and Samuel
  Welten.
\newblock Have a snack, pay with bitcoins.
\newblock In {\em IEEE P2P 2013 Proceedings}, pages 1--5. IEEE, 2013.

\bibitem{bei2011bounded}
Xiaohui Bei, Wei Chen, Shang-Hua Teng, Jialin Zhang, and Jiajie Zhu.
\newblock Bounded budget betweenness centrality game for strategic network
  formations.
\newblock {\em Theoretical Computer Science}, 412(52):7147--7168, 2011.

\bibitem{bergamini2018improving}
Elisabetta Bergamini, Pierluigi Crescenzi, Gianlorenzo D'angelo, Henning
  Meyerhenke, Lorenzo Severini, and Yllka Velaj.
\newblock Improving the betweenness centrality of a node by adding links.
\newblock {\em Journal of Experimental Algorithmics (JEA)}, 23:1--5, 2018.

\bibitem{decker2015fast}
Christian Decker and Roger Wattenhofer.
\newblock A fast and scalable payment network with bitcoin duplex micropayment
  channels.
\newblock In {\em Symposium on Self-Stabilizing Systems}, pages 3--18.
  Springer, 2015.

\bibitem{deng2002routing}
Hongmei Deng, Wei Li, and Dharma~P Agrawal.
\newblock Routing security in wireless ad hoc networks.
\newblock {\em IEEE Communications magazine}, 40(10):70--75, 2002.

\bibitem{dolev2010routing}
Shlomi Dolev, Yuval Elovici, and Rami Puzis.
\newblock Routing betweenness centrality.
\newblock {\em Journal of the ACM (JACM)}, 57(4):25, 2010.

\bibitem{gencer2018decentralization}
Adem~Efe Gencer, Soumya Basu, Ittay Eyal, Robbert Van~Renesse, and Emin~G{\"u}n
  Sirer.
\newblock Decentralization in bitcoin and ethereum networks.
\newblock {\em arXiv preprint arXiv:1801.03998}, 2018.

\bibitem{guha2005experimental}
Saikat Guha and Neil Daswani.
\newblock An experimental study of the skype peer-to-peer voip system.
\newblock Technical report, Cornell University, 2005.

\bibitem{heilman2015eclipse}
Ethan Heilman, Alison Kendler, Aviv Zohar, and Sharon Goldberg.
\newblock Eclipse attacks on bitcoin’s peer-to-peer network.
\newblock In {\em 24th $\{$USENIX$\}$ Security Symposium ($\{$USENIX$\}$
  Security 15)}, pages 129--144, 2015.

\bibitem{lightning2018paylittle}
Alyssa Hertig.
\newblock Coindesk: You can now get paid (a little) for using bitcoin’s
  lightning network, 2018.

\bibitem{kim2018measuring}
Seoung~Kyun Kim, Zane Ma, Siddharth Murali, Joshua Mason, Andrew Miller, and
  Michael Bailey.
\newblock Measuring ethereum network peers.
\newblock In {\em Proceedings of the Internet Measurement Conference 2018},
  pages 91--104. ACM, 2018.

\bibitem{koshy2014analysis}
Philip Koshy, Diana Koshy, and Patrick McDaniel.
\newblock An analysis of anonymity in bitcoin using p2p network traffic.
\newblock In {\em International Conference on Financial Cryptography and Data
  Security}, pages 469--485. Springer, 2014.

\bibitem{li2017survey}
Xiaoqi Li, Peng Jiang, Ting Chen, Xiapu Luo, and Qiaoyan Wen.
\newblock A survey on the security of blockchain systems.
\newblock {\em Future Generation Computer Systems}, 2017.

\bibitem{malavolta2019anonymous}
Giulio Malavolta, Pedro Moreno-Sanchez, Clara Schneidewind, Aniket Kate, and
  Matteo Maffei.
\newblock Anonymous multi-hop locks for blockchain scalability and
  interoperability.
\newblock In {\em NDSS}, 2019.

\bibitem{marcus2018low}
Yuval Marcus, Ethan Heilman, and Sharon Goldberg.
\newblock Low-resource eclipse attacks on ethereum's peer-to-peer network.
\newblock {\em IACR Cryptology ePrint Archive}, 2018:236, 2018.

\bibitem{nakamoto2008bitcoin}
Satoshi Nakamoto et~al.
\newblock Bitcoin: A peer-to-peer electronic cash system.
\newblock 2008.

\bibitem{naoumov2006exploiting}
Naoum Naoumov and Keith Ross.
\newblock Exploiting p2p systems for ddos attacks.
\newblock In {\em Proceedings of the 1st international conference on Scalable
  information systems}, page~47. ACM, 2006.

\bibitem{nemhauser1978analysis}
George~L Nemhauser, Laurence~A Wolsey, and Marshall~L Fisher.
\newblock An analysis of approximations for maximizing submodular set
  functions—i.
\newblock {\em Mathematical programming}, 14(1):265--294, 1978.

\bibitem{network2018cheap}
Raiden Network-Fast.
\newblock cheap, scalable token transfers for ethereum, 2018.

\bibitem{poon2016bitcoin}
Joseph Poon and Thaddeus Dryja.
\newblock The bitcoin lightning network: Scalable off-chain instant payments,
  2016.

\bibitem{rohrer2019discharged}
Elias Rohrer, Julian Malliaris, and Florian Tschorsch.
\newblock Discharged payment channels: Quantifying the lightning network's
  resilience to topology-based attacks.
\newblock {\em arXiv preprint arXiv:1904.10253}, 2019.

\bibitem{rosenfeld2014analysis}
Meni Rosenfeld.
\newblock Analysis of hashrate-based double spending.
\newblock {\em arXiv preprint arXiv:1402.2009}, 2014.

\bibitem{sompolinsky2013accelerating}
Yonatan Sompolinsky and Aviv Zohar.
\newblock Accelerating bitcoin's transaction processing. fast money grows on
  trees, not chains.
\newblock {\em IACR Cryptology ePrint Archive}, 2013(881), 2013.

\bibitem{trillo2013stress}
Manny Trillo.
\newblock Stress test prepares visanet for the most wonderful time of the year.
\newblock {\em URl: http://www. visa.
  com/blogarchives/us/2013/10/10/stress-testprepares-visanet-for-the-most-wonderful-time-of-the-year/index.
  html}, 2013.

\end{thebibliography}

%\section*{Backup: Fundamentals}\label{sec:model}

%This section presents our model of the Lightning network and the different approaches regarding the routing algorithms.

%We model the Lightning network using a directed graph, where the vertices are nodes, and each edge represents a one direction channel (bidirectional channels will be represented with two edges)\footnote{Note that in this model we do not distinguish between bidirectional channels and two directed opposite channels}. 

%The transactions in the Lightning network are private (by using onion routing) therefore we can not quantify the number of transactions the goes between the nodes. To perform an analysis of the transactions, we use the strong assumption that every pair of nodes execute a single transaction each way. Note that we will use different transaction sizes to examine different behaviors. Moreover, note that not every two nodes may execute a transaction, due to the path's length limit in the implementation (20 hops).

\balance 

\appendix
% \todo{Don't forget to delete it before submission}

% \section{Things to discuss and polish}

% \begin{itemize}
%     \item regarding title: both has a price: predictable and unpredictable! can we say that?
%          or something about topology/channel vs routing?
%     \item figures with less white space
%     \item captions with fullstop?
%     \item make dates consistent
%     \item got : obtained
%     \item did we introduce fuzz properly?
% \end{itemize}

% \section{Technical Stuff}

\begin{claim} \label{claim::solving_clique_game}
    The value of the game that is presented in Table~1 is $k \cdot \big( \frac{H}{\mid V \mid - 1} - I \big)$, and the Nash Equilibrium is that the attacker will play ``direct" with probability $\frac{k}{\mid V \mid - 1}$, and the defender with probability $\frac{1}{\mid V \mid - 1}$.
  \end{claim}
  \begin{IEEEproof}
    We will solve the game using the known fact: let $(\sigma_1, \sigma_2)$ be a Nash Equilibrium, and $U$ be the utility function of player 1. Then any pure strategy $s$ within the support of $\sigma_2$ holds the property that $U(\sigma_1, \sigma_2) = U(\sigma_1, s)$ and equals to the value of the game.
    
    Now getting back to the proof, assume that in Nash Equilibrium the defender choose ``direct" with probability $p$, then:
    $$ U((p, 1-p), direct) = p (-H -1) + (1-p)(-H \frac{k-1}{V - 2}-2) $$
    $$ U((p, 1-p), indirect) = p(-1) + (1-p)(-H\frac{k}{V - 2} - 2) $$
    from the fact above we get equality, therefore:
    $ -Hp -p -H\frac{k-1}{V - 2} -2 +pH\frac{k-1}{V - 2} + 2p = -p -H\frac{k}{V - 2} -2 +pH\frac{k}{V - 2} + 2p $ which is exactly
    $$pH(-1 + \frac{k-1}{V - 2}) - H\frac{k-1}{V - 2} = -H\frac{k}{V - 2} + pH\frac{K}{V - 2}$$ therefore $p = \frac{1}{V-1}$
    
    And the same for the attacker: if he will choose ``direct" with probability q then:
    $$ U_2(direct, (q, 1-q)) = q(H-Ik) + (1-q)(-Ik) $$
    $$ U_2(indirect, (q, 1-q)) = q (H \frac{k-1}{V - 2} - Ik) + (1-q)(H \frac{k}{V - 2} - Ik) $$
    Therefore: $qH -Ik = qH(\frac{k-1}{V - 2} - \frac{k}{V - 2}) + H\frac{k}{V - 2} -Ik$ which is exactly $qE = -q + k$ therefore $q = \frac{k}{V-1}$.
    
    Now, in order to get the value of the game for the attacker, we can simply calculate $U_2(pure\_direct, (q, 1-q)) = \frac{kH -Ik^2 -IEk -Ik + Ik^2}{V-1} = k\cdot(\frac{H}{V-1} - I)$
  \end{IEEEproof}

\end{document}